\documentclass[manuscript, nonacm]{acmart}

\usepackage{amsmath,amsfonts}
\usepackage{physics}
\usepackage{algorithmic}
\usepackage{graphicx}
\usepackage{textcomp}
\usepackage{xcolor}
\usepackage{tabularx}

\usepackage{multirow}

\newcommand{\circlenumber}[1]{\raisebox{.5pt}{\textcircled{\raisebox{-.9pt} {#1}}}}

\newcommand{\boldup}[1]{\textup{\textbf{#1}}}

\begin{document}

\title{qSIEVE: Efficient qLDPC Memory via Systolic Movement in Atom Arrays}



\author{Joshua Viszlai}
\affiliation{%
  \institution{The University of Chicago}
  \city{Chicago}
  \country{USA}}

\author{Willers Yang}
\affiliation{%
  \institution{The University of Chicago}
  \city{Chicago}
  \country{USA}}

\author{Sophia F. Lin}
\affiliation{%
  \institution{The University of Chicago}
  \city{Chicago}
  \country{USA}}
\affiliation{%
  \institution{AWS Center for Quantum Computing}
  \country{USA}}

\author{Junyu Liu}
\affiliation{%
  \institution{The University of Chicago}
  \city{Chicago}
  \country{USA}}
\affiliation{%
  \institution{The University of Pittsburgh}
  \city{Pittsburgh}
  \country{USA}}

\author{Natalia Nottingham}
\affiliation{%
  \institution{The University of Chicago}
  \city{Chicago}
  \country{USA}}

\author{Jonathan M. Baker}
\affiliation{%
  \institution{The University of Texas, Austin}
  \city{Austin}
  \country{USA}}

\author{Frederic T. Chong}
\affiliation{%
  \institution{The University of Chicago}
  \city{Chicago}
  \country{USA}}

\renewcommand{\shortauthors}{J. Viszlai, W. Yang, S. F. Lin, J. Liu, N. Nottingham, J. M. Baker, F. T. Chong}



\begin{abstract}

As quantum machines have scaled up in their number of qubits, significant research has turned towards increasing their fidelity with quantum error correction codes. Although promising results have been shown with the surface code, which only requires near-neighbor connections between qubits, the high qubit overhead of such local codes promises to be problematic. Consequently, recent work has explored non-local quantum LDPC (qLDPC) codes, which have good asymptotic encoding rates. Despite theoretical progress, hardware implementations of these codes has been a longstanding challenge.

At the experimental level, demonstrations of movement based communication on atom arrays suggest this is a powerful new primitive to achieve non-local connectivity. Leveraging this, we present a protocol for implementing non-local qLDPC codes in hardware. Our protocol, qSIEVE, is a co-design of such codes with movement in atom arrays. qSIEVE defines a restricted family of qLDPC codes that can be implemented efficiently with systolic movement.

We then quantify the utility of qSIEVE in the context of a complete fault tolerant architecture. We compare the cost of implementing benchmark programs in a standard, surface code only architecture and a mixed architecture where data is stored in qLDPC memory with qSIEVE and loaded to surface codes for computation. 

    
\end{abstract}


\maketitle

\section{Introduction}

In order to execute most quantum algorithms, we need orders of magnitude lower error rates than we can achieve physically. For example, chemistry and factoring algorithms can require error rates below $10^{-15}$ whereas reasonable expectations for physical error rates on many platforms are $10^{-3}$~\cite{beverland2022assessing}. To alleviate this gap between algorithmic requirements and physical capabilities, many researchers are focusing on developing fault-tolerant qubits. In a fault-tolerant quantum system, an application's quantum information is protected by underlying quantum error correcting (QEC) codes, allowing for continuous detection and correction of errors. In this setting the use of a logical quantum memory stores program qubits as logical qubits, each of which may be comprised of 10s-100s of physical qubits in a QEC code. 

To date, the theory of fault-tolerance has been well studied~\cite{calderbank1996good, shor1996fault, steane1996error, terhal2015quantum}; however, physical implementations of fault tolerance are nascent, requiring non-trivial adaptions of theory to hardware. Within this recently growing area of research, local codes such as the surface code~\cite{fowler2012surface} and color code~\cite{bombin2006topological} have been dominant~\cite{google2023suppressing, ryan2022implementing, krinner2022realizing, postler2022demonstration}, and for good reasons--codes only requiring local qubit connectivity are the most feasible for platforms limited to nearest-neighbor connections. Unfortunately, it's known that such local codes have poor encoding rates, introducing a $>$100x physical qubit overhead, and that non-local parity checks are necessary to produce QEC codes with better encoding rates~\cite{baspin2022quantifying}. 

\begin{figure}[t!]
    \centering
    \includegraphics[width=0.45\linewidth]{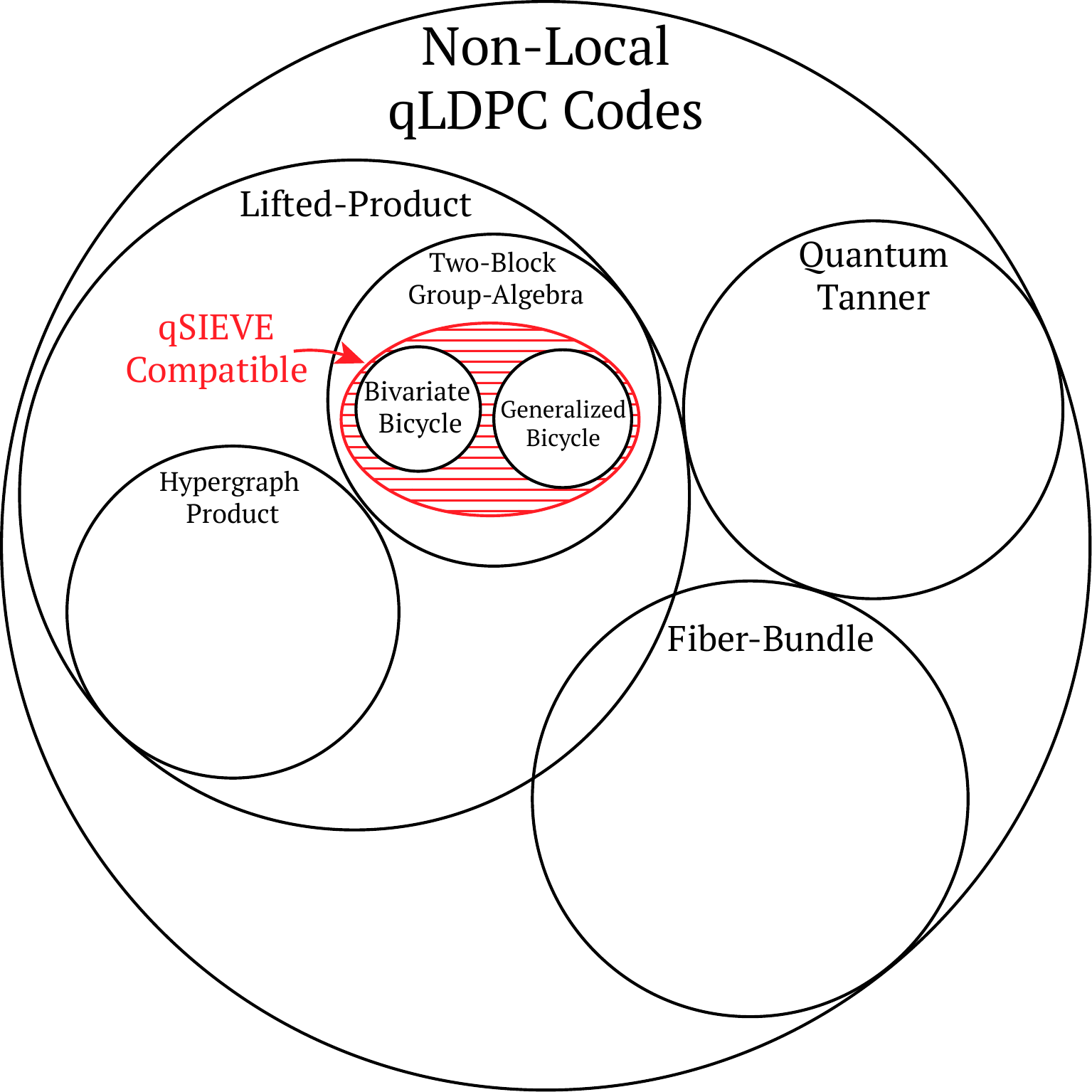}
    \caption{An overview of the large space of non-local qLDPC codes~\cite{ErrorCorrectionZoo}. We identify a restricted subset of codes in this space (highlighted in red) and present qSIEVE, a co-designed memory protocol using this restricted set of non-local codes with the available non-locality in reconfigurable atom arrays.
    }
    \label{fig:code-venn}
\end{figure}

Although superconducting devices have received a lot of attention, on-chip fabrication has constrained qubits to planar, nearest-neighbor connectivity, limiting a potential memory system's encoding rate. Recently, however, a breakthrough result~\cite{bluvstein2024logical} has brought considerable attention to quantum computers made from reconfigurable atom arrays. In these systems, qubits are suspended in free space via optical traps which can be moved in real-time, enabling non-local interactions not available in on-chip devices. This difference in innate connectivity indicates a different set of QEC codes should be used that takes advantage of the available non-locality to improve the encoding rate of a logical quantum memory.

In this work, we address this need and present qSIEVE, a protocol that identifies and enables a restricted set of non-local codes (Figure~\ref{fig:code-venn}) for quantum computers made from reconfigurable atom arrays. Codes within this set, such as generalized bicycle codes~\cite{kovalev2013quantum} and bivariate bicycle codes~\cite{bravyi2023high} have a reduced overhead of up to 10x compared to the surface code. qSIEVE's scheduler is also fast, enabling all checks to be measured in $\sim$3 ms, leading to measuring a round of stabilizers 5-11x faster than other protocols for non-local codes in atom arrays~\cite{xu2023constant}. We further validate qSIEVE through detailed circuit-level simulations and address different hardware control constraints. Importantly, qSIEVE enables multiple code blocks to be implemented with shared controls, which is extremely favorable for scaling to sizes needed for advantageous quantum algorithms. To this end, we propose and evaluate how qSIEVE can serve as a scalable quantum memory system.

We also quantify qSIEVE's utility in a full fault-tolerant architecture that includes logical operations. We analyze an architecture with three types of QEC systems. 1) Generalized-bicycle codes that serve as efficient quantum memory, leveraging their good encoding capabilities 2) A small number of surface codes that perform computation, leveraging their good logical capabilities, and 3) Ancilla systems that teleport data between qLDPC memory and surface code compute, which can be implemented in atom arrays~\cite{xu2023constant}. Treating this mixed code architecture as a simple memory hierarchy, we evaluate a suite of benchmark programs and compare with a surface code only architecture. Overall we find that for most programs of interest the spatial savings of generalized-bicycle codes outweighs the temporal overhead of loading and storing data between memory and compute.



\section{Background}
For background on quantum computing fundamentals we refer to~\cite{ding2020quantum, nielsen2001quantum, rieffel2011quantum}. In this paper we give relevant background on quantum error correction and atom array quantum computers.

\subsection{Quantum Error Correction}\label{sec:background_qec}
In this subsection we break down quantum error correcting codes into key components. To better convey the connections between QEC theory and systems-level features, each component is further separated into \circlenumber{1} theoretical underpinnings and \circlenumber{2} systems-level implications.

\boldup{Stabilizer Matrices:}
Stabilizers are pauli-product observables measured on \textbf{data qubits} in a QEC code. For example, a stabilizer $S=Z_1\otimes Z_2\otimes I_3$, measures the product of Z observables on qubits 1 and 2 but does not touch qubit 3. Groups of stabilizers are often described using \textbf{stabilizer matrices}, commonly referred to as parity check matrices in related literature. A stabilizer matrix, $H$, is a $m \times n$ binary matrix with rows indicating stabilizers and columns indicating data qubits. Entry $H_{i,j} \in \{0,1\}$ defines if data qubit $j$ is included in stabilizer $i$. To protect against both X and Z errors, and thus arbitrary errors on the logical qubit through linear combination, QEC codes have both X and Z stabilizers. In this work we only focus on codes where each stabilizer is uniformly Z-type or X-type. A single row $i$ in a Z-type stabilizer matrix, $H_Z$, therefore indicates a stabilizer measuring $Z$ on data qubits where $H_{Z,ij}=1$ and $I$ on data qubits where $H_{Z,ij}=0$. $H_X$ is defined equivalently for measuring stabilizers comprised of X observables. All stabilizers additionally need to commute. For a code with stabilizer matrices, $H_Z,H_X$, this enforces $H_XH_Z^T = 0 \mod 2$, that each pair of differing basis stabilizers has an even overlap. 

\emph{Systems-Level Implications:}
Stabilizers can be viewed synonymously as \textbf{parity checks} which allow indirect information to be gained about the underlying data qubits without fully collapsing our quantum state. These checks are the defining feature of a specific QEC code, and different constructions of the stabilizer matrices can define different families of codes. Although many constructions exist, in practice only a few unique codes may be needed to enable a fault-tolerant quantum system. For example, a single code used repeatedly for all program qubits can serve as a logical quantum memory.

\begin{figure}[H]
    \centering
    \includegraphics[width=0.55\linewidth]{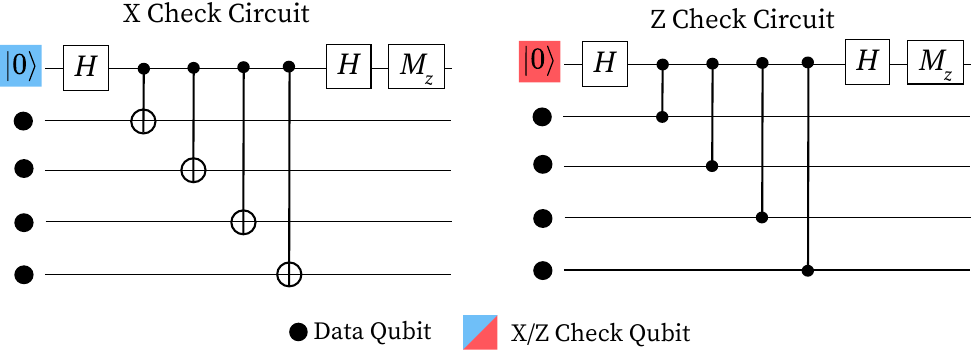}
    \caption{Circuits for performing an X parity check and a Z parity check.}
    \label{fig:check_circ}
\end{figure}

\boldup{Parity Check Circuits:}
Figure~\ref{fig:check_circ} shows examples of circuits which measure weight-4 parity checks. Each controlled gate introduces a +1 or -1 phase on the check qubit depending on the state of the data qubit in the checked basis. The resulting phase on the check qubit indicates the parity of the data qubits in the checked basis, which can then be measured out. In general, a weight $w$ parity check requires $w$ controlled operations, 1 for each constituent data qubit.

\emph{Systems-Level Implications:} 
Parity check circuits are a key connection between the system's physical and logical layers. Since these circuits need to be physically executed for all parity checks in our QEC code, they can introduce great stress on the connectivity of the device as each data qubit is involved in multiple parity checks. Furthermore, the time taken to measure all parity checks contributes to the \textbf{logical cycle time} of a code. Quality mapping of a code's qubits to a device is therefore critical to maximize circuit-level parallelism and minimize the logical cycle time. 

\boldup{Defining Logical Qubits}
Given the stabilizer matrices, $H_Z,H_X$, of a code with $n$ \textit{physical} data qubits, we can define $k$ unique \textit{logical} qubits where $k=n - \text{rk}(H_X) - \text{rk}(H_Z)$, corresponds to a $2^k$ dimensional logical Hilbert space within our $2^n$ dimensional physical Hilbert space. Independent logical observables are defined as:
\begin{align*}
    Z_{L} &\in \text{ker}(H_X)\setminus \text{rowsp}(H_Z) \\
    X_{L} &\in \text{ker}(H_Z)\setminus \text{rowsp}(H_X)
\end{align*}
and each logical qubit $i$ is made of an anticommuting pair of such observables:
\begin{align*}
    X_{L,i}Z_{L,j}^T + Z_{L,j}X_{L,i}^T = \delta_{i,j} \mod 2
\end{align*}
Note that the choice of which data qubits support which logical qubits is not necessarily unique, i.e. there may be many solutions for $X_{L,i}, Z_{L,i}$ that give rise to $k$ logical qubits. Furthermore, we can define the \textbf{code distance}, $d$, as the minimum weight possible observable $X_{L,i}, Z_{L,i}$. In general, finding the minimum code distance is NP-Hard, and so sampling-based approximations are often used instead~\cite{pryadko2023qdistrnd}. The parameters $[[n,k,d]]$ summarize a code's performance. For example, the rotated surface code has parameters $[[d^2, 1, d]]$

\emph{Systems-Level Implications:} The \textbf{encoding rate} $k/n$ indicates the qubit overhead incurred by a specific QEC code, and the code distance $d$ is the minimum number of physical errors that equate to a logical error. Codes with increased code distance have exponentially lower logical error rates, and so during compilation a code distance should be chosen to ensure the logical circuit meets fidelity requirements.

\boldup{Syndrome Decoding:}
We define a \textbf{syndrome}, $s$, as a $m \times 1$ binary vector where $s_i \in \{0,1\}$ indicates whether stabilizer $i$ changed. The \textbf{decoding} problem is to deduce the most likely physical error vector $e$ such that $He = s$. This is a classical problem that is NP-Hard to solve exactly~\cite{hsieh2011np}. In practice, decoding algorithms vary based on the specific QEC code being decoded. 

\emph{Systems-Level Implications:} A key constraint when decoding is performing it fast enough. If the rate of syndrome generation is higher than the rate of syndrome decoding, then a backlog occurs which exponentially slows down the computation~\cite{terhal2015quantum}. Broadly, real-time decoding is a classical problem that is its own subject of research which has seen lots of recent improvements~\cite{das2022afs, ravi2023better, vittal2023astrea}.

\subsection{Local vs. Non-local Codes}
\begin{figure}[t]
    \centering
    \includegraphics[width=0.45\linewidth]{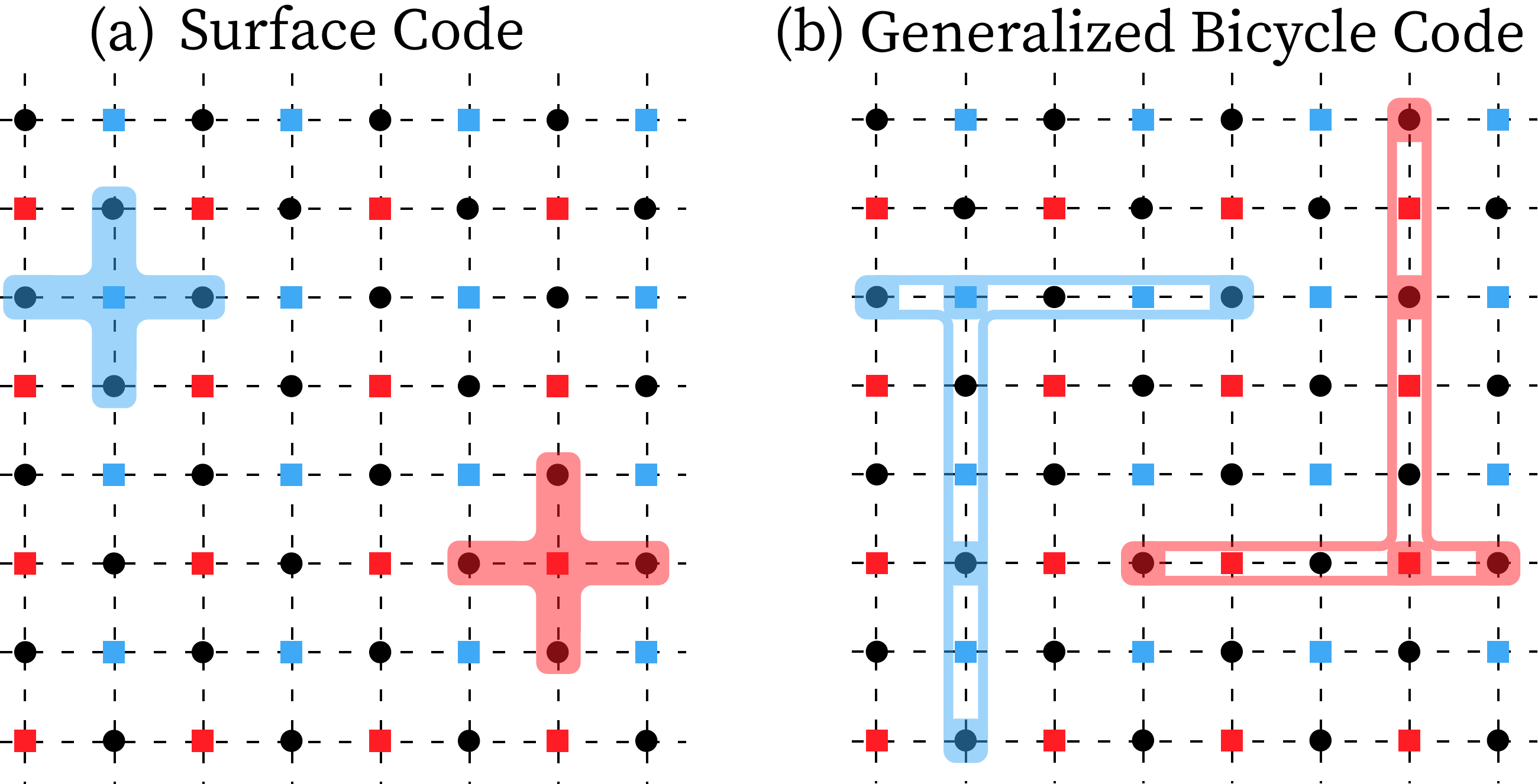}
    \caption{Check structure, indicating the data qubits (circles) involved in each parity check (squares). (a) The surface code (local). 
    (b) An example of a generalized bicycle code (non-local). The structure is identical for all checks of the same type, with blue indicating X-type checks and red indicating Z-type checks. 
    }
    \label{fig:code-comparison}
\end{figure}

We refer to~\cite{baspin2022quantifying} for a formal definition of non-locality. As an intuitive definition, we say a code is non-local if any mapping of the qubits to the device creates parity check qubits whose data qubits are far away, requiring long-range entangling operations. In the example shown in Figure~\ref{fig:code-comparison} the surface code is local because each parity check only needs nearest neighbor connections to interact with its data qubits. In the generalized-bicycle code, however, the check qubits have data qubits across the device, requiring long range operations. These long range operations are the key challenge for implementing non-local codes in hardware.

\subsection{Atom Arrays}
\boldup{Trapping and Movement:}
Atom arrays are quantum computers made from a 2D arrangement of optically trapped atoms~\cite{bluvstein2022quantum,jenkins2022ytterbium,ma2022universal,barnes2022assembly}. These devices typically use two types of trapping beams. A Spatial Light Modulator (SLM) creates \textit{static} traps for atoms, and 2D Acousto-Optic Deflectors (AODs) create \textit{reconfigurable} traps. Atoms can be \textit{transferred} from SLM traps to AOD traps, and the AOD traps can then be translated, stretched, or squeezed, enabling mid-circuit movement of 2D grids of qubits~\cite{bluvstein2022quantum}. These devices also have long coherence times on the order of seconds~\cite{barnes2022assembly, ma2022universal, jenkins2022ytterbium} and high-fidelity gates~\cite{evered2023high}.

\boldup{Two-Qubit Gates:}
Multi-qubit gates in these devices are mediated by the Rydberg interaction. The $|1\rangle$ state is coupled to a highly-excited Rydberg state, which blockades the same transition on neighboring atoms via van der Waals forces. The resulting interaction creates a CZ gate between atoms. By moving atoms that need to interact near each other, a laser pulse can be used to execute many CZ gates in parallel~\cite{evered2023high}. Furthermore, the strength of the Rydberg interaction scales $\propto \frac{1}{r^6}$ where $r$ is the distance from the excited atom. This means cross-talk can be removed by spacing non-interacting atoms sufficiently far apart.

\section{Motivation}
The first step in the path towards realizing a fault-tolerant quantum system is encoding program qubits into logical qubits via some QEC code. Such a system can be viewed as a logical quantum memory, and although many QEC codes exist in theory~\cite{ErrorCorrectionZoo}, implementations for most of them in actual hardware are still unknown. Most codes that we do know how to implement in hardware are geometrically local codes and suffer from notoriously poor encoding rates, leading to costs in the millions of physical qubits for large-scale quantum algorithms~\cite{beverland2022assessing}. 

This cost in space is the key challenge to be addressed for quantum memory. It's known that for many algorithms of interest, a large portion of a quantum circuit consists of \textit{idle volume}~\cite{litinski2022active}, meaning most of a program's qubits will be sitting idly at any given time. This is the task of quantum memory and so minimizing its physical footprint will have a large impact on the overall circuit cost. Additionally the quantum memory does not need to be responsible for holding qubits during active steps of a quantum program. Theoretical constructions exist that can be used to move qubits between codes~\cite{cohen2022low}, and furthermore it's known how to perform such constructions in atom arrays and with the codes qSIEVE targets~\cite{xu2023constant, bravyi2023high}. 

Reducing the space costs associated with quantum memory requires the use of better codes. Many such codes are often categorized as qLDPC codes, and require some amount of non-local operations. The challenge now becomes adapting theoretical constructions of these qLDPC codes to existing hardware platforms. This is an area with few existing solutions (detailed in Section~\ref{sec:related}). Since this is tightly connected to the hardware of interest, there are even fewer solutions for atom arrays. Solving this problem and identifying a competitive memory protocol for atom arrays is therefore the primary motivation of this work.

\section{Related Work}~\label{sec:related}
Xu et al.~\cite{xu2023constant} proposed an implementation of hypergraph-product and lifted-product codes on atom arrays. They implement parity check circuits via 1D atom rearrangements that are applied to rows and columns of an atom array, matching the product structure of the codes they consider with the product structure of AODs. In this work, qSIEVE targets a different set of codes where we find parity checks can be implemented by moving check qubits collectively in 2D. Table~\ref{tab:movement_compare} summarizes the key differences between their implementation and qSIEVE. In addition to measuring a round of stabilizers 5-11x faster, the codes implementable by qSIEVE also have better encoding rates at small sizes. For example, at $d=6$, codes compatible with qSIEVE have a $\sim4\times$ higher encoding rate.
 
There has also been prior work on implementing other types of QEC codes, such as the surface code~\cite{auger2017blueprint}, on atom arrays. Compiling algorithms for atom arrays in the Noisy Intermediate-Scale Quantum (NISQ) era has also been a subject of research~\cite{baker2021exploiting,patel2022geyser,tan2022qubit,nottingham2023decomposing,litteken2022reducing,patel2023graphine, wang24atomique, tan2024dpaa, tan2024enola}, however many of these techniques would require adaptation to apply in a fault-tolerant setting.

Some prior works explored how non-local qLDPC codes can be implemented on other types of hardware~\cite{bravyi2023high, tremblay2022constant, pattison2023hierarchical}. Notably, IBM's recent work~\cite{bravyi2023high} showed that the Tanner graphs of bivariate-bicycle codes can be partitioned into two planar subgraphs. As a consequence, the codes can be realized on planar superconducting chips with long range links. 



\section{qSIEVE: Efficient Quantum Memory in Atom Arrays}\label{sec:gb_memory}
Here we introduce qSIEVE: an efficient memory protocol that implements generalized-bicycle~\cite{kovalev2013quantum} and bivariate bicycle~\cite{bravyi2023high} codes in atom array quantum computers. For simplicity we refer to both families of codes as generalized-bicyle codes going forward, using the construction we detail in Section~\ref{gb:construct}.  



\begin{figure*}
    \centering
    \includegraphics[width=0.9\textwidth]{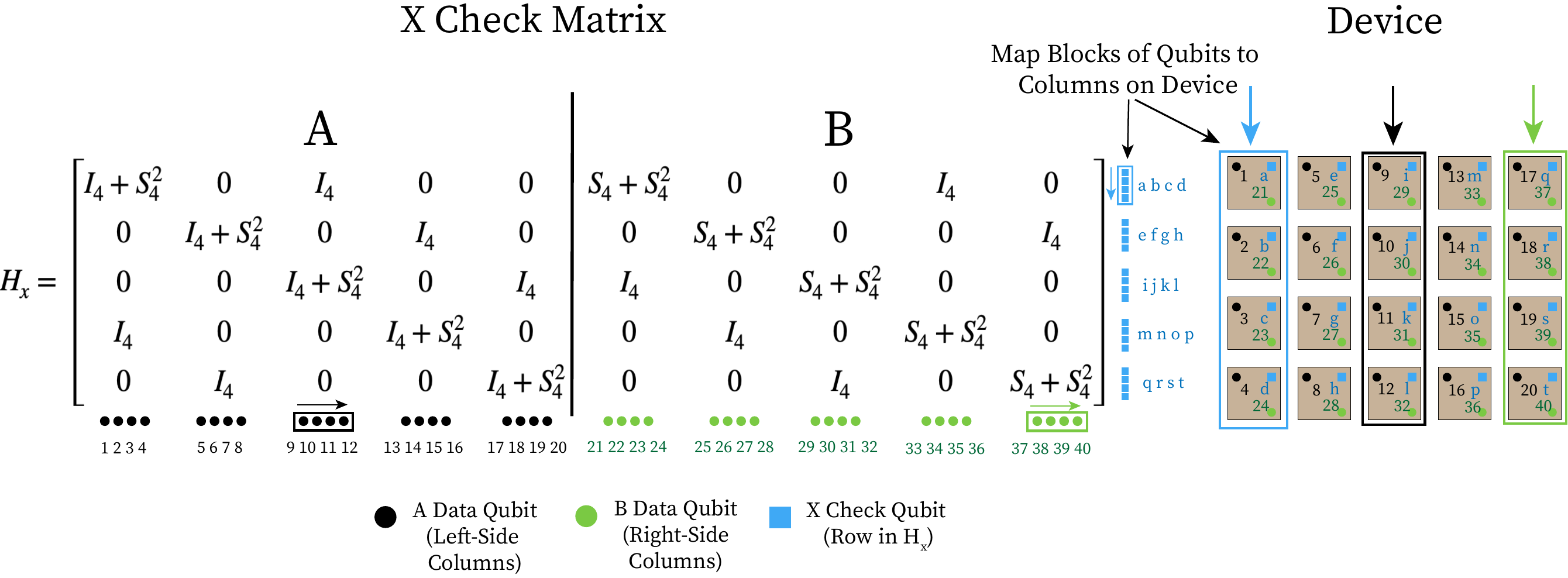}
    \caption{Mapping of qubits from an example X check matrix ($H_x$) to columns on the device. This check matrix is derived from parameters $l=5, m=4, a(x,y)=1+y^2+x^2, b(x,y)=y+y^2+x^3$. The mapping from the Z check matrix ($H_z$) to device columns is done in a corresponding manner, with Z check qubits located in the bottom left corner of each group of four qubits.}\label{fig:mat-to-device}
\end{figure*}

\subsection{Generalized-Bicycle Codes}\label{gb:construct}
As a family of qLDPC codes, generalized bicycle codes have promising asymptotic encoding rates. Additionally, recent work has found many instances of generalized-bicycle codes with good encoding rates, even at small sizes~\cite{bravyi2023high, lin2023quantum}. 
Here we give the code construction of generalized-bicycle codes used in qSIEVE. 

\boldup{Permutation Matrices:}
The check matrices of the generalized bicycle codes are created using a collection of permutation matrices that identify data qubits (columns) with parity checks (rows). Take for example,
\begin{equation*}
    S_3 = 
    \begin{bmatrix}
        0 & 1 & 0 \\
        0 & 0 & 1 \\
        1 & 0 & 0
    \end{bmatrix};
    S_3^2 = 
    \begin{bmatrix}
        0 & 0 & 1 \\
        1 & 0 & 0 \\
        0 & 1 & 0
    \end{bmatrix};
    S_3^3 = S_3^0= 
    \begin{bmatrix}
        1 & 0 & 0 \\
        0 & 1 & 0 \\
        0 & 0 & 1
    \end{bmatrix}
\end{equation*}
Each power $S_3^{p}$ maps row $i$ to column $i + p \mod 3$. In general we will define this type of matrix as $S_l^p$ for some size $l$.

As parity checks are performed on a collection of data qubits, we can sum many distinct permutation matrices together to create a matrix $A$. For example, 
\begin{equation*}
    A = S_3 + S_3^2 = 
    \begin{bmatrix}
        0 & 1 & 1 \\
        1 & 0 & 1 \\
        1 & 1 & 0
    \end{bmatrix}
\end{equation*}
defines 3 parity checks, each on two data qubits.

\boldup{Generalized-Bicycle Code Check Matrices:}
As generalized bicycle codes are a two-block quantum code\cite{kovalev2013quantum} rather than a single matrix $A$ as described above, we use two such matrices $A,B$ that are constructed independently. The code's X and Z check matrices are then defined as:
\begin{equation}\label{eq:stabilizers}
    H_x = (A | B), 
    H_z = (B^T | A^T)
\end{equation}
When A and B are each constructed from summing permutation matrices $S_l^p$ this check matrix construction corresponds to the original definition of generalized-bicycle codes introduced in~\cite{kovalev2013quantum}.

However, recent work\cite{lin2023quantum, bravyi2023high} has looked at a more general construction of generalized-bicycle codes where $S_l^p$ is replaced by a tensor product of two independent permutation matrices $S_l^p \otimes S_m^q$.

For convenience, it is often easier to write $S_l^p$ as $x^p$ and $S_m^q$ as $y^q$ where
\begin{equation}\label{eq:stab_term}
    x^p y^q = S_l^p\otimes S_m^q
\end{equation}
Then our matrices $A$ and $B$ are expressed succinctly as polynomials over $x$ and $y$. As an example, with $l=5, m=4$ $A$ and $B$ could be:
\begin{align*}
    A = 1 + y^2 + x^2 = I_5 \otimes I_4 + I_5 \otimes S_4^2 + S_5^2 \otimes I_4 \\
    B = y + y^2 + x^3 = I_5 \otimes S_4 + I_5 \otimes S_4^2 + S_5^3 \otimes I_4
\end{align*}
which when combined into the check matrices as $H_x = (A|B);$ $H_z = (B^T | A^T)$ defines a specific generalized-bicycle code with $lm=20$ X parity checks, 20 Z parity checks, and 40 data qubits. Where each parity check is on 6 data qubits (3 defined by $A$ and 3 defined by $B$).

\boldup{Parameters of the Construction:}
Here we summarize the construction by highlighting key parameters:
\begin{itemize}
\item $l,m$: Together define the size of the X and Z check matrices of the code.
\item $x^py^q$: A single permutation-like matrix defined in Equation~\ref{eq:stab_term} which identifies each data qubit with a parity check.
\item $a(x,y), b(x,y)$: Identifies data qubits to parity checks. The total weight of both polynomials together defines the weight of each parity check.
\end{itemize}

\boldup{[[n,k,d]] of Constructed Code:}
Given a specific choice of $l,m,a(x,y),b(x,y)$ we can derive $[[n,k,d]]$ for the constructed code.
\begin{itemize}
\item $n$: The number of data qubits, $n$, are the number of columns in $H_x$ and $H_z$ which is $2*l*m$. 
\item $k$: The number of encoded qubits, $k$, can be calculated as $k = 2 * l * m - \text{rk}(H_X) - \text{rk}(H_Z)$.
\item $d$: As mentioned in Section~\ref{sec:background_qec} the code distance, $d$, is NP-Hard to calculate in general. For small codes $d$ can be calculated exhaustively, however, most of the time sampling-based approximations are used.
\end{itemize}

\begin{figure*}
    \centering
    \includegraphics[width=0.7\linewidth]{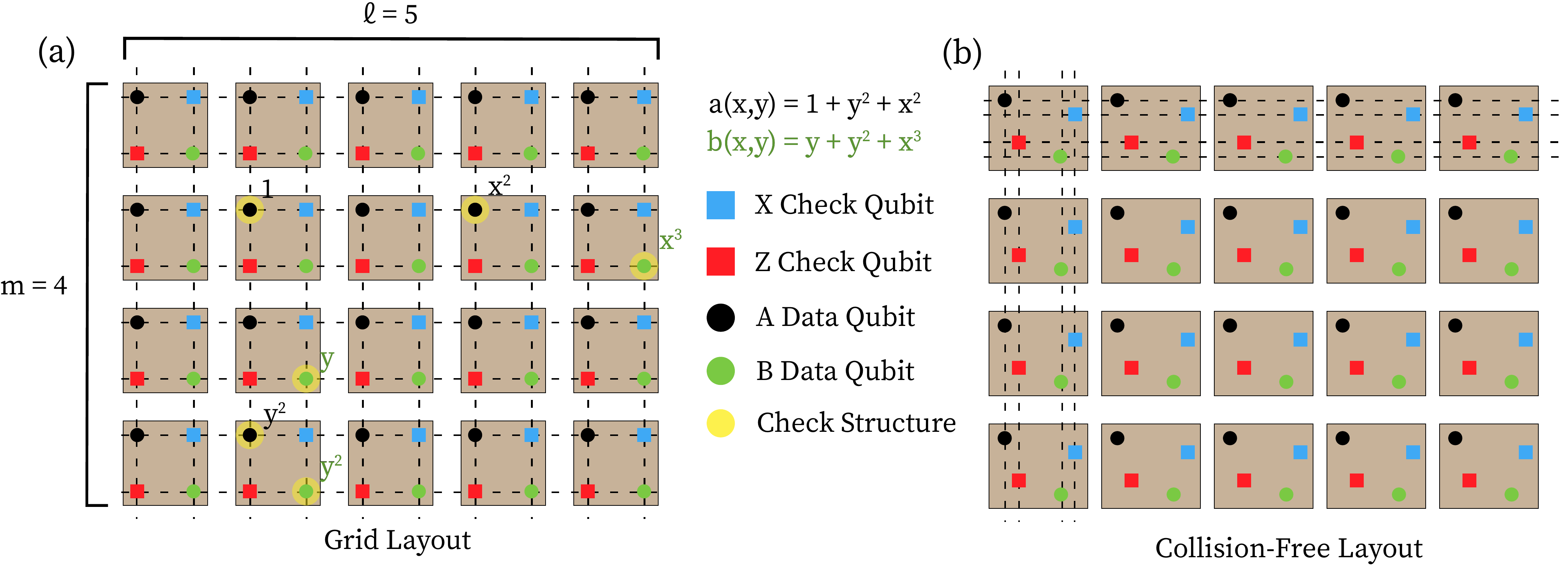}
    \caption{(a) A continuation of the example in Figure~\ref{fig:mat-to-device} showing the complete qubit layout we use for generalized-bicycle codes. 
    Each subgrid (X check, Z check, A data, B data) is mapped identically and interleaved. The highlighted check structure shows the relative position of data qubits from a given $X$ check qubit. This structure is identical for all $X$ check qubits up to periodic boundaries. $Z$ check qubits have a mirrored structure. (b) A modified version of the layout that simplifies collision-free movement of X and Z check qubits.} 
    \label{fig:gb-layout}
\end{figure*}

\begin{table*}[t]
    \centering
    \vspace{1em}
    \begin{tabular}{c|c|c|c|c|c|c}
        $[[n,k,d]]$ & $l,m$ & $a(x,y)$ & $b(x,y)$ & Check Weight & \# Steps/Op & Source \\
        \hline
        $[[72,12,6]]$ & $6,6$ & $y+y^2+x^3$ & $y^3+x+x^2$ & 6 & 2 &~\cite{bravyi2023high}\\
        $[[90,8,10]]$ & $15,3$ & $y+y^2+x^9$ & $1+x^2+x^7$ & 6 & 2 &~\cite{bravyi2023high}\\
        $[[144,12,12]]$ & $12,6$ & $y+y^2+x^3$ & $y^3+x+x^2$ & 6 & 2 &~\cite{bravyi2023high}\\
        $[[128, 16, 8]]$ & $8,8$ & $y+y^2+y^5+x^6$ & $y^2+x^2+x^3+x^7$ & 8 & 2 & This work\\
        $[[72, 8, 10]]$ & $36,1$ & $1+x^9+x^{28}+x^{31}$ & $1+x+x^{21}+x^{34}$ & 8 & 2 & ~\cite{lin2023quantum}\\
        $[[96, 10, 12]]$ & $12,4$ & $1+y+xy+x^9$ & $1+x^2+x^7+x^9y^2$ & 8 & 2 or 4 & ~\cite{lin2023quantum}\\
    \end{tabular}
    \caption{Selected codes used in numerical simulations}
    \label{tab:code_selection}
\end{table*}

\subsection{Qubit Mapping}\label{gb:layout}

Figure~\ref{fig:mat-to-device} gives an example picture of how qSIEVE maps qubits in the check matrices onto the device.  Motivated by the tensor product structure $S_l^p \otimes S_m^q$ we map all qubits to a position $(row,col)$ on a $m\times l$ grid. We then have four types of qubits to map to the device: X check qubits (rows of $H_x$), Z check qubits (rows of $H_z$), A data qubits (columns of $A$), and B data qubits (columns of $B$). Each of these is represented by a total of $lm$ rows/columns. For each type we identify qubit $i$ with the $i^{\text{th}}$ row/column and then map to the device at position $(row,col)=(i\mod m, \lfloor i/m \rfloor)$.

Figure~\ref{fig:gb-layout} expands on the example and shows the full layout as well as how the check structure translates to the mapping. Figure~\ref{fig:gb-layout}b also introduces a small modification to the mapping that simplifies fast movement. The collision-free layout allows X and Z check qubits to be moved purely vertically and horizontally without overlapping with a stationary data qubit, which would cause a collision. 


\subsection{Systolic Check Operations}\label{parity_struct}
We first note that after our mapping, terms $x^py^q$ define a highly regular, parallelizable operation which we define as a \textbf{systolic check operation}.
Each term $x^py^q$ defines a mapping from checks to data qubits with the relative position between the two being identical for all X(Z) checks*. This regular structure is ideal for AOD movement, and our router exploits the available parallelism for fast cycle times. 

\boldup{*Handling Periodic Boundaries:}
Due to the presence of permutation matrices in the code construction, the offset dictated by systolic check operation $x^py^q$ from a check qubit to one of its data qubits may go beyond the bounds of the layout. In such a case, the bounds of the layout are `periodic' and the offset wraps around to the opposite side. To formalize this behavior, we derive all possible relative positions between check and data qubits in our mapping. With no periodicity, the relative position between check $i$ and its data qubit is $(row,col)=(q,p)$. If $i \geq m(l-p)$, we have horizontal periodicity and the column offset is changed to $(p - l)$. If $i\mod m \geq (m - q)$, we have vertical periodicity and the row offset is changed to $(q - m)$. Overall, this gives four possible relative positions: $(q,p), (q, p-l), (q-m, p), (q-m,p-l)$. Additionally, if $p=0$ or $q=0$, this reduces to only two possibilities, which is the case for most of the codes we simulate. Only one code we simulate has mixed terms ($p,q>0$).


         

\subsection{Circuit Routing}
To implement generalized-bicycle codes, the circuits of Figure~\ref{fig:check_circ} need to be executed for all X and Z checks in the code. Since single qubit rotations can be performed with high fidelity in atom arrays~\cite{levine2022dispersive}, we focus on routing the long-range two qubit gates between check and data qubits. These long-range gates are mediated by atom movement using an AOD, therefore the router should aim to minimize this movement cost. 

\boldup{Traveling Salesman Formulation:}
Figure~\ref{fig:qsieve-tsp} gives an overview of how qSIEVE's restricted code selection and mapping procedure allows for the routing problem to be treated as a traveling salesman problem. We note that X and Z checks are routed separately; CZ gates for all X checks are performed first, followed by CZ gates for all Z checks. Furthermore, all X (Z) check qubits are moved collectively during routing. In addition to reducing movement time, doing so minimizes the number of times atoms need to be transferred between movable, AOD traps and static, SLM traps, which can be a source of error. The stops are derived from the relative positions described in Section~\ref{parity_struct} for a code's systolic check operations. Conveniently, if the space beyond the boundaries is unoccupied (described further in Section~\ref{sec:tiled}) then moving to stop $(q,p)$ leaves all check qubits with a different, periodic offset beyond the boundary. CZ gates can then be performed only between the intended check-data qubit pairs. Importantly, the reverse also holds. When moving to the stop $(q,p-l)$ the check qubits that were excluded at stop $(q,p)$ are now within the boundaries and can perform their CZ gates. This can be seen in the case of $x^2$ shown in Figure~\ref{fig:qsieve-tsp}. To solve for the minimal routing, we use the Concorde TSP solver. Fortunately, by moving X (Z) check qubits collectively, the number of stops scales with the check weight, not the code size. And since we expect the check weight to be small to minimize error propagation, qSIEVE can scalably route generalized-bicycle codes of increasing size.

\begin{figure*}
    \centering
    \includegraphics[width=\textwidth]{Assets/qsieve-overview.pdf}
    \caption{Overview of mapping and routing in qSIEVE. Given code parameters for a generalized-bicycle code, qubits are mapped to the device to allow for systolic check operations. Routing is then formulated as a traveling salesman problem and solved using the Concorde TSP solver. The routing in the figure shows the path for X check qubits. A mirrored path is used for Z check qubits. The two example stops explain how the $x^2$ operation is eventually performed for all checks. The time step on the left performs all CZ gates for check qubits with offset $(0, 2)$, and the time step on the right performs all CZ gates for check qubits with offset $(0, -3)$.}
    \label{fig:qsieve-tsp}
\end{figure*}


  


\subsection{Movement Costs}\label{gb:movement}
We assume spacing between atom sites of 5$\mu m$ and acceleration of 0.02$\mu m$/$\mu s^2$, and we model movement costs according to~\cite{xu2023constant}, with all movement done along the Manhattan distance between two consecutive stops. The movement time associated with a given check is thus, 
\begin{equation}
\left(\sqrt{6\cdot\Delta x / 0.02} + \sqrt{6\cdot\Delta y / 0.02}\right)
\end{equation}
where $\Delta x$ is the distance that needs to be moved along the horizontal direction, and $\Delta y$ is the distance along the vertical direction. By using the collision-free grid arrangement shown in Fig.~\ref{fig:gb-layout}b, we ensure that no collisions occur during movement. We summarize our movement costs in Table~\ref{tab:movement_compare}.

\begin{table}[]
    \centering
    \begin{tabular}{c|c|c|c}
          Code &  Encoding Rate & Round of Checks & Tool \\
         \hline
         $[[72,12,6]]$ & 0.166 & 2.71 ms & qSIEVE \\
         $[[90,8,10]]$ & 0.088 & 2.97 ms & qSIEVE \\
         $[[144,12,12]]$ & 0.083 & 2.97 ms  & qSIEVE \\

         $[[128,16,8]]$ & 0.125 & 3.77 ms & qSIEVE \\
         $[[72,8,10]]$ & 0.111 & 5.25 ms & qSIEVE \\
         $[[96,10,12]]$ & 0.104 & 4.52 ms & qSIEVE \\
         
         $[[625,25,6]]$ & 0.04 & 14.85 ms & \cite{xu2023constant} \\
         $[[2500, 100, 12]]$ & 0.04 & 20.1 ms & \cite{xu2023constant} \\
         $[[544, 80, \leq 12]]$ & 0.147 & 33.4 ms & \cite{xu2023constant} \\
         \hline
         $[[121,1,11]]$ & 0.008 & $\sim$ 1 ms & \cite{bluvstein2022quantum} \\
         ($d=11$ surface code) & & & 
    \end{tabular}
    \caption{Summary of qSIEVE Performance}
    \label{tab:movement_compare}
\end{table}

\subsection{Hardware Compatibility}\label{gb:hardware}
A key hardware feature required for our implementation is the collective movement of 2D grids of check qubits. In atom arrays, coherent movement of a 2D array of qubits via an AOD has been experimentally demonstrated in the context of quantum error correction~\cite{bluvstein2022quantum,xu2023constant,bluvstein2024logical}. High-fidelity parallel CZ gates on adjacent qubits and single-qubit rotations have also been demonstrated experimentally with error rates of $\sim5\times 10^{-3}$ and $<10^{-3}$, respectively~\cite{evered2023high, ma2023high, levine2022dispersive}. Most notably, these error rates are below the threshold for many QEC codes, including the surface code and the generalized-bicycle codes we consider. Various methodologies for mid-circuit measurement also exist in experiment~\cite{singh2022dual, graham2023mid, shea2020submillisecond, bluvstein2024logical}.

The scale needed for a single memory block with qSIEVE has also been demonstrated as machines with $>$ 250 atoms exist in experiment~\cite{ebadi2021quantum, singh2022dual, bluvstein2024logical}. For reference, the smallest code we simulate only requires 144 atoms (using the $[[72,12,6]]$ code).

\section{Simulating Memory Performance}\label{gb:simulate}
We evaluate the performance of our proposed implementation via numerical simulations. We construct generalized-bicycle codes from code parameters $l, m, a(x,y), b(x,y)$ as described in Section~\ref{sec:gb_memory} and compile their parity check circuits into a Stim~\cite{gidney2021stim} circuit. 

\subsection{Error Modeling}~\label{gb:error}
In our simulations we use a full circuit-level noise model for a physical error rate $p$ that includes propagation of errors during the parity check circuits. Single qubit gates are followed by $\{X,Y,Z\}$ errors with probabilities $\frac{p}{3}$. Two qubit gates are followed by $\{I,X,Y,Z\}\otimes\{I,X,Y,Z\}\setminus \{I\otimes I\}$ errors with probabilities $\frac{p}{15}$. Qubit reset and measurement are flipped with probability $p$. 

In our compilation process we also model the underlying movement operations. Idle errors are added after each movement operation using pauli twirling approximations~\cite{ghosh2012surface}. The movement time is calculated as described in Subsection~\ref{gb:movement}.

\subsection{Decoding}~\label{gb:decoding}
Including initialization, our simulations decode over 
a standard $d$ parity check cycles. For decoding, we use the BP-OSD decoder~\cite{panteleev2021degenerate} available in the ldpc python package~\cite{roffe_decoding_2020, Roffe_LDPC_Python_tools_2022}. The decoding hyperparameters we use are a maximum number of iterations of 10,000, the min-sum BP method, and the combination-sweep OSD with order 10. As is standard for CSS codes, we separate circuit-level decoding into X and Z sub-problems, each only decoding errors that flip X or Z checks, respectively. 

\begin{figure}
    \centering
    \includegraphics[width=0.5\linewidth]{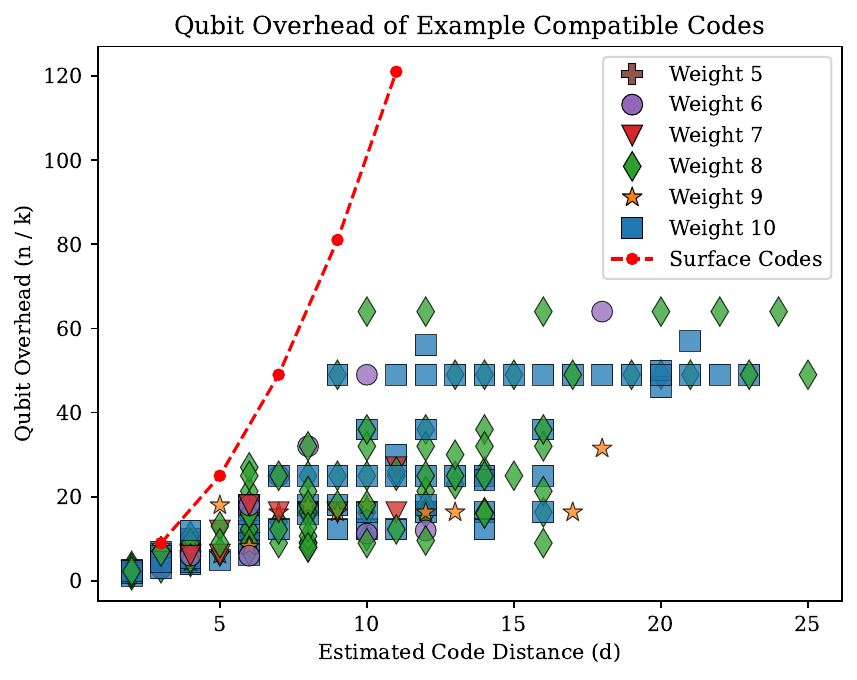}
    \caption{Overhead of physical qubits per logical qubit (lower is better) for a collection of qSIEVE-compatible codes with $n < 150$. Estimated distance is an upper bound derived from a sampling-based approximation.}
    \label{fig:example-codes}
\end{figure}

\subsection{Code Selection}
As shown in Figure~\ref{fig:example-codes}, the set of codes compatible with qSIEVE is quite large. We find codes ourselves by sampling parameters $l, m, a(x,y), b(x,y)$, and we select codes from recent works that can be described as generalized-bicycle codes. However rather than simulate all possible codes, to evaluate the performance of qSIEVE we select a representative set, shown in Table~\ref{tab:code_selection}. \# Steps/Op indicates the number of movements and gate pulses necessary for each check operation as described in Section~\ref{parity_struct}. Mixed $x,y$ terms require 4 steps due to the presence of both horizontal and vertical periodicity. All other terms require 2 steps. Three of the codes we choose are from recent work on bivariate bicycle codes~\cite{bravyi2023high}. These can be seen as instances of generalized-bicycle codes with trinomials $a(x,y)$ and $b(x,y)$ where terms are only either $x^p$ or $y^q$. 
We also choose two codes from recent work on two-block group algebra codes~\cite{lin2023quantum}. In this context the generalized-bicycle codes we address are a limiting case when the group being considered is a cyclic group. 


\subsection{Results}
\begin{figure}[t]
    \centering
    \includegraphics[width=\linewidth]{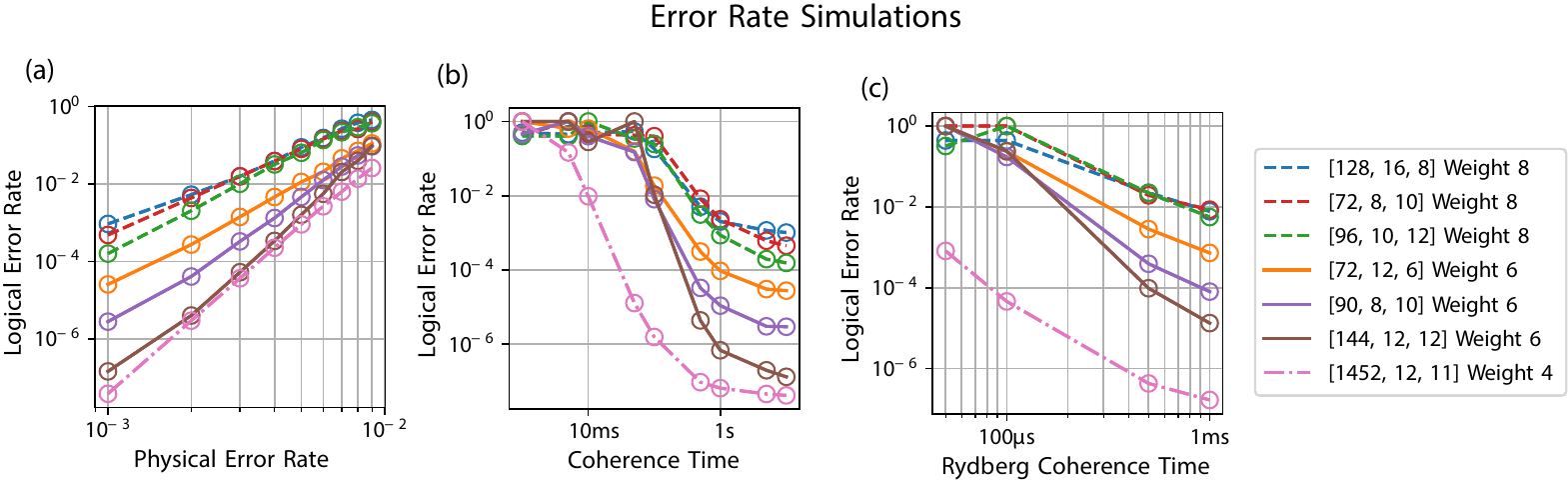}
    \caption{Logical error rate simulations of the codes in Table~\ref{tab:code_selection} using qSIEVE. The simulations also include a surface code plot for comparison, generated from simulating 12 surface codes ($[[1452, 12, 11]]$).}
    \label{fig:all-err_rates}
\end{figure}
\boldup{Memory Performance:}
Figure~\ref{fig:all-err_rates}a shows the results of our circuit-level simulations. The logical observables for each encoded qubit were found using the same methodology described in Section 8.1 of ~\cite{bravyi2023high}. The physical error rate corresponds to the parameter $p$ in our error modeling described in Section~\ref{gb:error} with coherence times set to 10 seconds. We simulate generalized-bicycle codes and the surface code for $d$ rounds. For each data point, we collected $N_s$ shots and defined $N_e$ as the number of shots where a logical error occurred. $N_s$ was chosen dynamically until at least 1,000 errors occurred, or $10^9$ shots occurred. All data points accumulated at least 100 errors. The plotted logical error rate $p_L$ is then defined per round as $p_L = 1 - (1 - (N_e/N_s))^{1/d} \approx \frac{(N_e/N_s)}{d}$. 

Our results indicate that our implementation of the weight 6 generalized-bicycle codes has a comparable logical error rate to surface codes but with a significantly better encoding rate. We also find the weight 8 codes have poorer performance. This can be attributed to the longer circuits necessary for the higher weight checks leading to increased error propagation. Future work could improve this by using techniques such as flag fault-tolerance~\cite{chamberland2018flag}.

\boldup{Coherence Sensitivity:}
Figure~\ref{fig:all-err_rates}b shows the result of our sensitivity study to the coherence time of the qubits. In this study, $p$ is set to $10^{-3}$ and the coherence time is varied. Logical error rate $p_L$ is again defined per round as $p_L = 1 - (1 - (N_e/N_s))^{1/d} \approx \frac{(N_e/N_s)}{d}$. We find that the generalized-bicycle codes are more sensitive to coherence times than the surface code, as expected. However, the generalized-bicycle codes are still feasible since coherence times in experiment are on the order of seconds~\cite{barnes2022assembly, ma2022universal, jenkins2022ytterbium}.

\subsection{Effects of Global Control}
In this work we assume two-qubit gates can be localized such that only pairs of adjacent check and data qubits are addressed. Some experiments, however, employ alternative control schemes where two-qubit gate addressing is global within a two-qubit gate zone~\cite{bluvstein2024logical}. In the absence of noise this creates no difference--a Rydberg pulse on an idle qubit with no adjacent qubits results in an identity operation. But in reality an idle qubit excited to the Rydberg state may still experience decoherence errors proportional to the Rydberg state's coherence time. To evaluate the impact of this on qSIEVE when constrained to global addressing, we add in decoherence errors on idle qubits after two-qubit gates in our simulations. The decoherence time is modeled as the two-qubit gate time, which we estimate as 262 ns based on recent experiments~\cite{evered2023high}. The results are shown in Figure~\ref{fig:all-err_rates}c.

Our results serve as a conservative assessment of Rydberg decoherence's impact on logical error rates. We do not employ any mitigation strategies such as dynamical decoupling or erasure conversion~\cite{wu2022erasure} which would require a physical experiment. Reported Rydberg coherence times in experiment are $<100\mu$s, and so we conclude a naive application of qSIEVE in a global control scheme may not be viable in the near-term. 

\begin{figure}[t]
    \centering
    \includegraphics[width=0.5\linewidth]{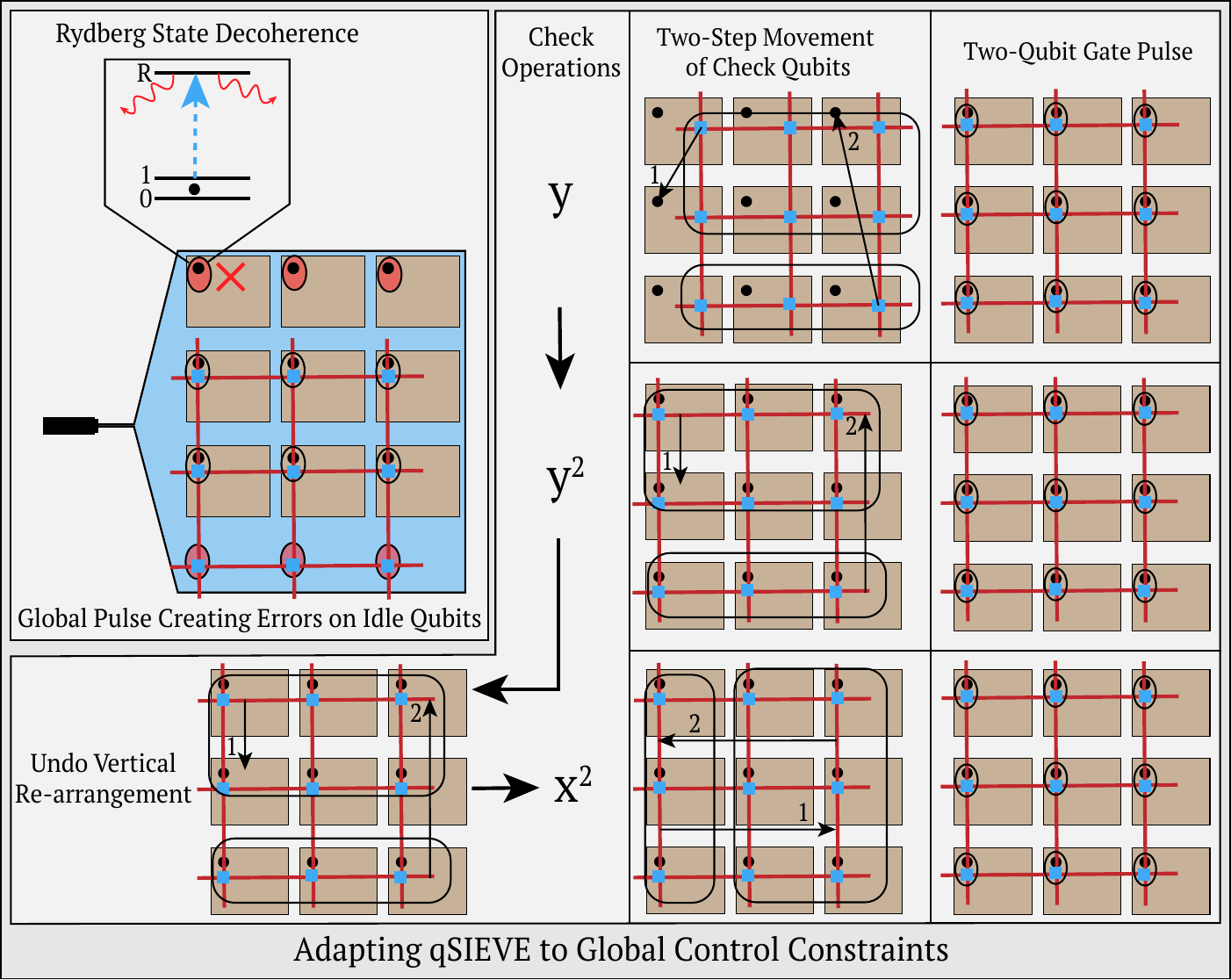}
    \caption{Modified qSIEVE protocol to avoid idle qubits in a zoned/global control scheme.}
    \label{fig:control-constraints}
\end{figure}

\boldup{Adapting to Zoned Controls:}
In Figure~\ref{fig:control-constraints} we motivate how qSIEVE can be adapted to a zoned control scheme, like the one employed in~\cite{bluvstein2024logical}. Given a pair of check and data subgrids (X, Z, A, B) in the gate zone, idle qubits can be omitted in each systolic check operation by both shifting the non-periodic qubits and the periodic check qubits before applying a two-qubit gate. This needs to be broken into two movement steps since crossing movement paths would cause undesirable heating and loss of atoms~\cite{bluvstein2022quantum}. Minor optimizations can be employed, however. For example, shown in Figure~\ref{fig:control-constraints}, ordering $y^2$ after $y$ means previously periodic qubits rearranged during $y$ can be moved with the non-periodic qubits in $y^2$. We report estimated movement costs in Table~\ref{tab:zone-move} based on a fully implemented protocol with localized control costs from Table~\ref{tab:movement_compare} reported again for comparison. We assume a separation of $20 \mu m$ between the storage and gate zones, and we find movement times are still low enough that we can regain the logical error rates reported in Figure~\ref{fig:all-err_rates}a and Figure~\ref{fig:all-err_rates}b in a zoned control system.

\begin{table}[]
    \centering
    \begin{tabular}{c|c|c}
    Code & Round (Localized) & Round (Zoned) \\
    \hline
    $[[72,12,6]]$ & 2.71 ms & 4.00 ms\\
    $[[90,8,10]]$ & 2.97 ms & 4.27 ms\\
    $[[144,12,12]]$ & 2.97 ms & 4.39 ms\\
    \end{tabular}
    \caption{Movement Times for Different Control Schemes}
    \label{tab:zone-move}
\end{table}

\section{Scaled Quantum Memory with qSIEVE}

In sections~\ref{sec:gb_memory} and~\ref{gb:simulate} we described the qSIEVE protocol and evaluated its performance for a single code instance, which we will refer to as a \textbf{memory block}. The capacity of a single memory block is limited to $k$, which for the codes we simulate is $< 20$, much less than the size of advantageous quantum algorithms. In this section we therefore examine qSIEVE in a quantum memory consisting of multiple memory blocks.

\subsection{Tiled Memory Designs}\label{sec:tiled}

Here, we evaluate a 2D tiled quantum memory where each block is an identical code implementable by qSIEVE as described in Section~\ref{sec:gb_memory}. We first focus on the control costs associated with increasing the number of memory blocks to meet program demand.

\boldup{Shared AOD:}
Since each memory block uses the same code instance, they all have identical systolic check operations. This means we can treat each block-level operation as a memory-level operation, using a single shared movement and gate schedule for the entire memory. From a control perspective, this means using the same AOD for all memory blocks in parallel, only requiring extra laser power for the new trapping beams. 

\begin{figure}[t]
    \centering
    \includegraphics[width=\linewidth]{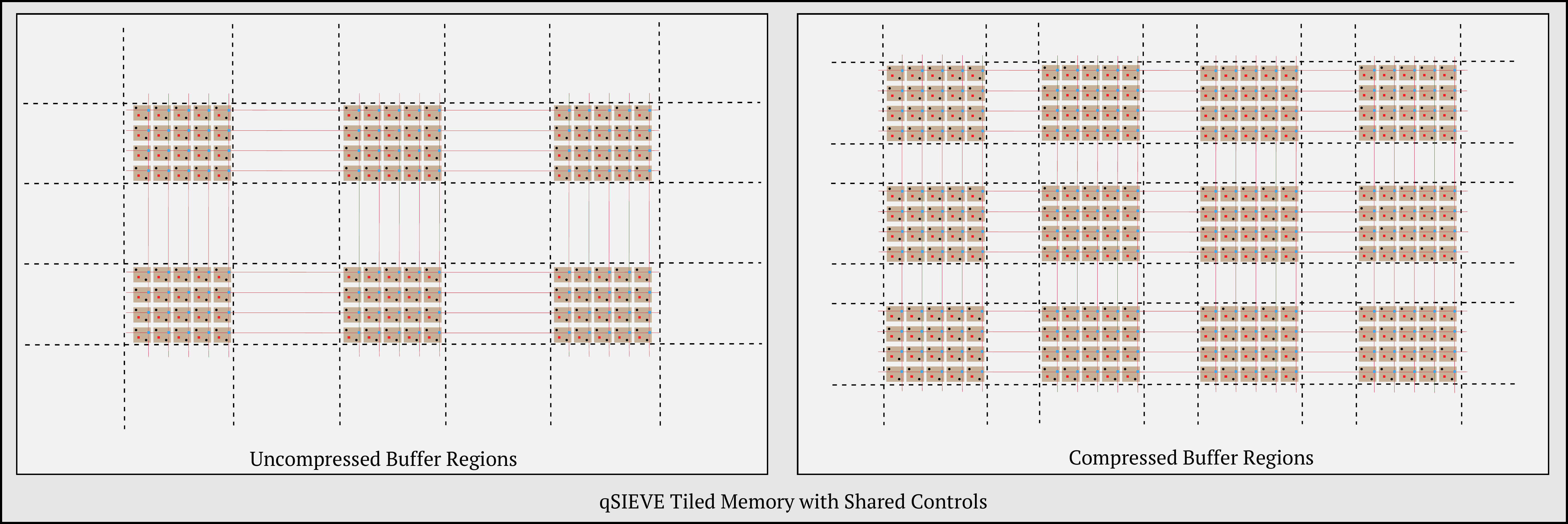}
    \caption{Tiled memory using qSIEVE with uncompressed (left) and compressed (right) buffer regions to prevent check qubits from bleeding into neighboring blocks during movement.}
    \label{fig:tiled-memory}
\end{figure}

\boldup{Shared Buffer Regions:}
Figure~\ref{fig:tiled-memory} depicts a tiled quantum memory using qSIEVE. Because movement may result in check qubits temporarily beyond the boundary of their block, we need to allocate buffer space to prevent these check qubits from bleeding into neighboring blocks. It will never be the case that two blocks need to simultaneously use buffer space in between them, as that would require them to have different movement paths. Instead, since movement is identical for all blocks, buffer regions can be shared, reducing the overall footprint.

\boldup{Compressing Buffer Regions:}
We can further reduce the footprint of the buffer regions by noting that the original spacing between check qubits shown in Figure~\ref{fig:gb-layout} is $2r$, where $r$ is the initial atom spacing. We assume $r$ is chosen as the distance at which no cross-talk occurs between idle atoms, and so to ensure no cross-talk occurs between check qubits in the buffer regions, we only need a spacing between them of $r$. This can be achieved smoothly with AOD movement, which supports stretching and squeezing of traps in addition to translation, so long as no moving AOD traps cross~\cite{bluvstein2022quantum}. As a result, we can reduce the size of buffer regions by modifying the movement path such that check qubits are squeezed from a spacing of $2r$ to $r$ after going beyond their block's boundaries.

\subsection{Memory Footprint}
\begin{figure}[t]
    \centering
    \includegraphics[width=\linewidth]{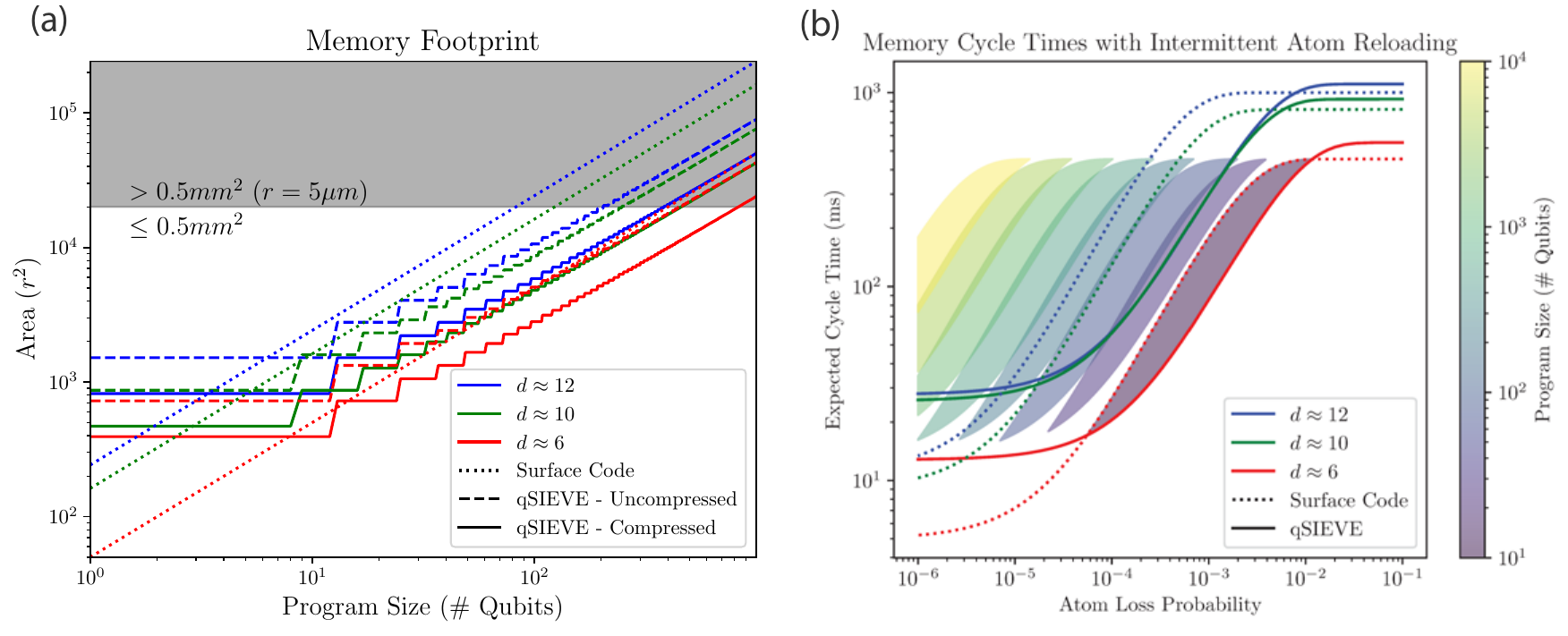}
    \caption{(a) Total memory footprint of qSIEVE vs. Surface Code. Area is expressed in terms of the atom spacing ($r$), which is assumed as $5 \mu m$ throughout this work. (b) Expected cycle time when accounting for cycles experiencing atom loss that require a 90 ms reloading step. Plotted lines are for 10 program qubits, and shaded regions indicate where qSIEVE is quicker than the surface code in the $d \approx 6$ case for varying program sizes. In general, increasing the number of program qubits shifts the curves to the left.}
    \label{fig:mem-footprint}
\end{figure}

For large-scale atom arrays, current experimental demonstrations of traps with $6,100$ atoms have practical upper limits around $1.5mm$ on the field of view diameter of the trapping beams~\cite{manetsch2024tweezer}, as such the physical footprint of our logical memory is important in addition to its physical qubit count. Figure~\ref{fig:mem-footprint}a shows the memory footprint when using qSIEVE versus surface codes for varying program sizes. Area is expressed in units of the atom spacing, $r$, which we assume is $5\mu m$ throughout this work based on experimental demonstrations~\cite{bluvstein2024logical, singh2022dual}. In spite of the need for buffer space, we find the higher encoding rate of the codes enabled by qSIEVE leads to a lower overall footprint when compared to surface codes. This advantage is furthered by the use of compressed buffer regions. As a useful point of interest, we also plot an area cutoff of $0.5mm^2$ when the atom spacing is $5 \mu m$. In the case of the $[[144,12,12]]$ code with qSIEVE using compressed buffer regions, this is about the area cost corresponding to 10,000 physical qubits, which is an expected upper bound on the capabilities of a single atom array processor~\cite{bluvstein2024logical}.

\subsection{Effects of Atom Loss}
In reconfigurable atom arrays, atoms may fall out of optical SLM or AOD traps, an event termed \textbf{atom loss}. Although atom loss probabilities are $<0.1\%$~\cite{bluvstein2022quantum}, we expect to run a quantum memory for millions of cycles in which case atom loss events are inevitable. 

\boldup{Logical Effects of Atom Loss:}
Detecting atom loss can be achieved via standard imaging of the atoms which can occur at the end of a round of parity checks during measurement. After loss has been detected, a recovery operation is needed to reload new atoms. This can be treated as an erasure error where atoms are reloaded into completely mixed states and the resulting effect on the QEC code is an equally probable $I,X,Y,\text{ or }Z$ error on each replaced qubit~\cite{wu2022erasure}. This type of error is easier to decode than the standard depolarizing errors we modeled in Section~\ref{gb:simulate}. If we are in a regime where $p_{loss} \ll p$, with $p_{loss}$ being the loss probability and $p$ being the underlying physical error rate, then we can expect no degradation of logical performance as the Pauli errors introduced during erasure conversion would be limited by $p_{loss}$ and dwarfed by those introduced by $p$. However, we note that detailed simulations of higher probability erasure errors in the non-local codes we consider have not been evaluated yet and may perform differently than results for local codes~\cite{wu2022erasure}.

\boldup{Reloading Delay:}
Recovering from atom loss requires reloading fresh atoms into the vacated traps. Real-time implementations of this in experiment have been achieved in $\sim 90 ms$~\cite{singh2023mid}. Notably, this reloading occurs on a larger timescale than our underlying logical cycles. If atom loss events are sufficiently likely then the effective cycle time can become much slower due to the need for reloading between rounds of parity checks. To evaluate the impact of this, we estimate the expected cycle time:
\begin{equation*}
    \mathbb{E}[t_{\text{cycle}}] = d\mathbb{E}[t_{\text{round}}] = d(t_{\text{checks}} + (1 - (1 - p_{\text{loss}})^{N})t_{\text{reload}})
\end{equation*}
where $d$ is the number of rounds in a logical cycle and $N$ is the total number of physical qubits. $t_{\text{checks}}$ is the standard time for a round of parity checks reported in Table~\ref{tab:movement_compare}. We assume loss is detected and reloading is employed after each round of parity checks.

Figure~\ref{fig:mem-footprint}b shows the expected logical cycle time for varying atom loss probabilities. We also shade the areas where a $d\approx6$ qSIEVE has a lower expected cycle time than the equivalent surface codes for varying program sizes. In general the curves shift to the left as program size increases. In our calculations, we assume a round of surface code checks takes $1 ms$~\cite{bluvstein2022quantum}. Interestingly, we find that in expectation, for a range of loss probabilities the increased number of physical qubits in surface code memory creates notably more rounds with atom loss leading to a higher cycle time compared to qSIEVE.

\section{Evaluation of Quantum Memory Hierarchy}\label{sec:ft}

\begin{figure}[t]
    \centering
    \includegraphics[width=0.5\linewidth]{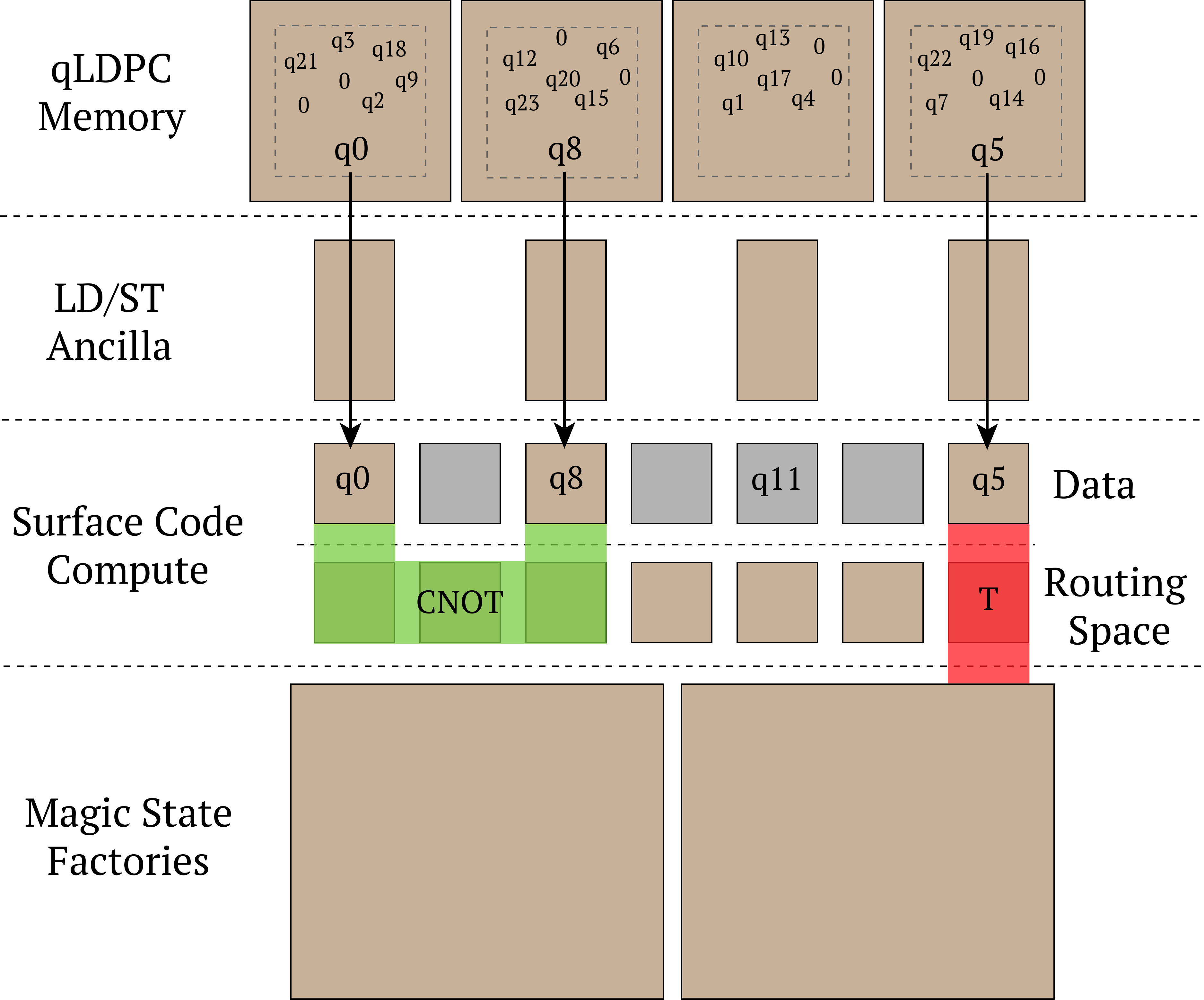}
    \caption{A planar layout of the hierarchical qLDPC+surface code architecture discussed in Section~\ref{sec:ft}.}
    \label{fig:ft-arch}
\end{figure}
In this section, we evaluate the utility of a proof-of-concept quantum memory hierarchy. We compare this against a fault-tolerant architecture based solely on surface codes~\cite{litinski2019game}.

Our evaluation of this quantum memory hierarchy serves as empirical evidence of the advantage of carefully designed hardware implementations of QEC memory, which is a crucial assumption justifying several recent works\cite{breuckmann2021quantum, bravyi2023high,xu2023constant}.

First, we describe details of the hierarchical architecture in Subsection~\ref{ssec:ft-details}. We consider an architecture, shown in Figure~\ref{fig:ft-arch}, where generalized-bicycle codes are used as efficient memory systems, surface codes are used for computation, and data transfer between the two is enabled by teleportation systems. This is a well-motivated architecture to consider as it also minimizes the hardware complexity needed to implement logic, and recent work~\cite{xu2023constant} suggests the LD/ST ancilla, which can be viewed as hypergraph-product codes~\cite{cohen2022low}, may be feasible to implement in atom arrays.  
We aim to show that, with well-motivated cost metrics, a hierarchical architecture with efficient quantum memory can outperform a surface code-only architecture on a wide range of benchmark programs. In our evaluation, we define cost as spacetime volume (qubit-seconds). This is a natural choice as qLDPC memory reduces space costs due to higher encoding rates while increasing time costs due to data movement. The spacetime volume therefore allows for an evaluation of this trade-off.

Based on spacetime volume, the relative cost of the two architectures is dependent on the program being run. We develop a compiler in Subsection~\ref{ssec:ft-compiler} and evaluate key benchmark circuits, matching program features such as serialization and T consumption to features of the hierarchical architecture through sensitivity studies described in Subsection~\ref{ssec:ft-results}.
Although our analysis is focused on generalized-bicycle codes in atom arrays, many insights, such as the impact of load/store time and the breakdown of costs per benchmark, have broad implications for future hierarchical fault-tolerant systems with a dedicated quantum memory component.

\begin{table}[t]
    \centering
    \begin{tabular}{c|c|c}
        Gate & \# Logical Cycles & Code Type\\
        \hline
        X, Y, Z, H, S & 0 & - \\
        CNOT & 2 & Surface Code\\
        T & 2 & Surface Code \\
        Load & 2 & LD/ST Ancilla\\
        Store & 1 & LD/ST Ancilla\\
    \end{tabular}
    \caption{Gate costs used in estimates}
    \label{tab:gate_costs}
\end{table}
\subsection{Hierarchical Architecture Details}\label{ssec:ft-details}
Figure~\ref{fig:ft-arch} shows the layout we consider for the hierarchical qLDPC+surface code architecture. In our evaluation we consider implementation costs for atom array quantum computers by default.

\textup{\textbf{Quantum Memory:}}
We assume all memory blocks are implemented with the same generalized-bicycle code using our protocol. This means the check structure is identical for all of memory, allowing all blocks to use the same AOD in parallel as described in Section~\ref{sec:tiled}. 

\textup{\textbf{LD/ST:}} We model LD/ST ancilla constructed according to~\cite{cohen2022low} to implement a $ZZ$ measurement between a qLDPC qubit and a surface code qubit. This can be treated as a hypergraph product code with a space cost of $\sim 2d^2$ qubits. This allows for the use of the circuit in Figure 3 of~\cite{poulsen2017fault}. Since the temporal cost of loads and stores is high we find this ancilla system best minimizes the overall spacetime volume by requiring only a single $ZZ$ operation.

LD/ST ancilla systems are laid out to connect memory blocks to a subset of the surface codes. We assume the use of an AOD allows for the LD/ST ancilla to selectively connect one of potentially many surface codes to the memory block. In the example of Figure~\ref{fig:ft-arch}, each LD/ST ancilla is assigned 1 memory block and 2 surface codes, except for the right-most which is assigned 1 surface code. We also make a simplifying assumption that each LD/ST ancilla can only operate on one qubit, meaning two logical qubits can't be loaded or stored from the same memory block simultaneously. 

\textup{\textbf{Surface Code Compute:}} We assume the surface codes use the wide design from~\cite{litinski2018lattice} and cost $2d^2$ qubits each, allowing single-qubit Clifford gates to be done in software and access to both $X$ and $Z$ boundaries.

The surface code compute section consists of a row of surface codes which house data, and a row of surface codes for routing CNOT and T gates via lattice surgery. Additionally, we can remove the routing surface codes if using movement-based transversal CNOTs. This is also the same layout used in the baseline surface code only architecture. 

\textup{\textbf{T State Factories:}} We assume a section of the device is dedicated to producing magic states in magic state factories. The underlying code is also surface codes, and we assume these surface codes connect to the routing space, allowing for T state injection. The T state factories we consider are described in~\cite{litinski2019magic}.

\subsection{~Cost Modeling}
\textup{\textbf{Gate Costs:}} 
Table~\ref{tab:gate_costs} shows the number of logical cycles we assume each operation takes to execute. CNOT requires a $ZZ$ measurement followed by a $XX$ measurement which takes 2 logical cycles when implemented using lattice surgery~\cite{horsman2012surface, litinski2019game} or 1 logical cycle when implemented using a transversal CNOT~\cite{bluvstein2022quantum}. T gates are performed via state injection using a CNOT gate, consuming a T state produced in a magic state factory. Load and Store both only require a $ZZ$ measurement between a qLDPC and surface code qubit, which takes 1 logical cycle. This is followed by measuring out the old qubit location. For Store this is the surface code, and can be done in a single measurement. For Load this is a qubit in qLDPC memory. To avoid disturbing the other qubits in the memory block, we must re-use the qubits in the LD/ST ancilla system to measure out the qLDPC qubit using a system according to~\cite{cohen2022low}. Like the LD/ST system, this can also be treated as a hypergraph product code.
In total, this means Load takes two logical cycles, one for measuring $ZZ$ and one for measuring out the old qLDPC qubit.

\textup{\textbf{LD/ST and Surface Code Costs:}} By default, our simulations assume that for the LD/ST ancilla, a round of measuring checks takes 2.5ms based on~\cite{xu2023constant}. The sensitivity of this assumption is evaluated in Figure~\ref{fig:sensitivity}. For surface codes, we assume one round of parity checks takes 1ms based on~\cite{bluvstein2022quantum}. For both codes, we assume a logical cycle is a standard $d$ rounds of parity checks. 

\textup{\textbf{T Factory Costs:}} The T factory is chosen to ensure the output T gate fidelity is low enough to reach overall program fidelity requirements. We model candidate T factories and production rates based on~\cite{litinski2019magic}. We also model potential stalls in benchmark programs due to in-progress T state production.

\begin{table*}[t]
    \centering
    \vspace{1em}
    \begin{tabular}{l|l|l|l|l|l}
         Benchmark Name&  Abbr.&Application& \# Qubits & Serialization& T Consumption \\
         \hline
         Quantum Adder \cite{li2023qasmbench} &  adder &Factoring& 28, 64, 118&High& Low\\
         Berstein-Vazarani \cite{li2023qasmbench,berstein1997bv} &  bv &Algorithm&30, 70, 140& High&  None\\
 Quantum Counterfeit Coin \cite{li2023qasmbench,terhal1998cc}& cc &Algorithm& 32, 64, 151& Low& None\\
         GHZ State Synthesis \cite{li2023qasmbench}&  ghz &State Prep.& 40, 78, 127 & High&  None\\
 Hubbard Prepare Oracle \cite{cirq_developers_2023_8161252, babbush2018lineart}& hb\_prep &State Prep.& 40, 70, 90, 100&Medium&  Low\\
         Hubbard Select Oracle \cite{cirq_developers_2023_8161252, babbush2018lineart}&  hb\_sel &Chemistry& 37, 76, 100, 162&High&   Medium\\
 Ising Model Simulation\cite{li2023qasmbench}& ising &Chemistry& 34, 66, 98 &Low& High\\
 Multiple-Control-Toffoli \cite{cirq_developers_2023_8161252, low2018trading}& mcx &Factoring& 30, 48, 120&High& Low\\
 Quantum Read Only Memory\cite{cirq_developers_2023_8161252, babbush2018lineart}& qrom &Chemistry& 18, 52, 125&Medium& Medium\\
 W State Synthesis\cite{li2023qasmbench}& wstate &State Prep.& 36, 76&High& High\\    \end{tabular}
    \vspace{0.2em}
    \caption{High-level descriptions of the benchmark circuits selected}
    \label{tab:benchcircs}
\end{table*}

\subsection{Compilation Methodology}\label{ssec:ft-compiler}
For compilation, we operate on input programs that have already been synthesized into a fault-tolerant gate set of Clifford + $T$ using standard approaches~\cite{ross2014optimal, nielsen2001quantum}. The main task of compilation is therefore to: 1) map qubits to qLDPC memory blocks, 2) insert load and store operations, and 3) route gates and T states in the surface code. We develop a proof-of-concept compiler with several reasonable heuristics to take advantage of program features such as serialization under a memory hierarchy but leave it as future work to conduct an in-depth study of compilation targeting the quantum memory hierarchy.

\textup{\textbf{Mapping Qubits:}} We use a graph-coloring mapper to assign program qubits to memory blocks. A slowdown that can be attributed to poor mapping occurs when a CNOT operates on two qubits in the same memory block. Since two qubits cannot be loaded from the same memory block simultaneously, their loads must be serialized. A reasonable heuristic therefore is to maximize the number of CNOT gates with qubits mapped to different memory blocks. We build an interference-style graph where qubits are nodes and edges indicate two qubits have a CNOT between them. Each memory block is treated as a color and the mapper aims to find a graph coloring that maximizes the number of edges with nodes of different colors. Coloring is then performed using a Chaitin-style approach~\cite{chaitin1982register}.

\textup{\textbf{Inserting Loads and Stores:}} 
We insert loads and stores greedily. For each time step in the program, we iterate through all operations in that time step.  If the support of the operation is already in the surface code, no LD/ST is needed. Otherwise, we schedule load operations when possible; that is, when there are free surface codes to load the support to and free LD/ST ancillae to perform teleportation. The operation is delayed otherwise. We say a qubit is active when there's an operation to be scheduled on it, and an inactive qubit may become active in future steps. Before executing a round of computation, we attempt to schedule additional LD/ST operations in anticipation of future usage, in a round of \textit{prefetching}. We first schedule load operations, if possible, on the qubit in each memory block that becomes active the earliest; then, we schedule store operations, if possible, on qubits in surface codes, \textit{if}  another qubit in the memory will become active sooner. To break ties, we'll prioritize one support of a CNOT gate when the other support is already in the surface code. These heuristics also implicitly optimize for \textit{qubit reuse}, since no qubits may become active sooner than a currently active qubit, meaning that an active qubit in the surface code will not be stored until all operations on it are executed. The only exception is a tiebreak when it becomes necessary to make space for a CNOT gate to preserve dependency.

\textup{\textbf{Routing Operations:}}
The compiler needs to route two types of gates in the surface code: CNOT and $T$. Routing is done greedily at each time step. To implement a CNOT gate via lattice surgery, a path of ancilla in an additional routing space is allocated between the two qubits, and for $T$ gates, a single routing ancilla between the data and the T state factories is allocated. If routing space cannot be allocated for an operation, the operation is delayed. In the movement-based implementation, no additional routing space is needed; however, we make a simplifying assumption that CNOT gates and T gates must be serialized since AOD-based movement cannot cross over itself.

\textup{\textbf{Ensuring Program Fidelity:}}
To ensure the compiled program meets fidelity requirements, we profile the circuit and choose requisite code distances and magic state factories. Overall program fidelity is estimated as\begin{align*}
    f_{\text{prog}} &= (1-\epsilon_{\text{mem}})^{n_{\text{blocks}} n_{\text{cycles}}}\\
    &\times(1-\epsilon_{\text{ldst}})^{n_{\text{blocks}}n_{\text{ldst}}}\\
    &\times(1-\epsilon_{\text{surface}})^{n_{\text{surface}}n_{\text{cycles}}}\\
    &\times(1-\epsilon_{\text{rz}})^{n_{\text{rz}}}\\
    &\times(1-\epsilon_{\text{t}})^{n_{\text{t}}}
\end{align*}
$\epsilon_{\text{mem}}$ is the logical error rate of a memory block, $\epsilon_{\text{ldst}}$ is the logical error rate of a LD/ST ancilla, $\epsilon_{\text{surface}}$ is the logical error rate of a surface code, $\epsilon_{\text{rz}}$ is the approximation error of RZ gates via discrete T sequences, and $\epsilon_{\text{t}}$ is the logical error rate of a produced T state. $n_{\text{blocks}}$ is the number of memory blocks which is also the number of LD/ST ancilla, $n_{\text{ldst}}$ is the number of LD/ST gates, $n_{\text{surface}}$ is the number of surface codes, $n_{\text{cycles}}$ is the compiled program length, $n_{\text{rz}}$ is the original number of RZ gates, and $n_{\text{t}}$ is the number of T gates in the synthesized program. Logical error rates are defined per cycle and derived from simulations, as described in Section~\ref{gb:simulate}, and T state error rates are taken from~\cite{litinski2019magic}. A few benchmarks required a larger code distance memory block than we simulated. To accommodate this we chose the $[288,12,18]$ code from~\cite{bravyi2023high} which is implementable with our protocol. We use a conservative logical error rate estimate of $10^{-9}$ for the physical error rate of $10^{-3}$.

\begin{figure}[t!]
    \centering
    \includegraphics[width=0.55\linewidth]{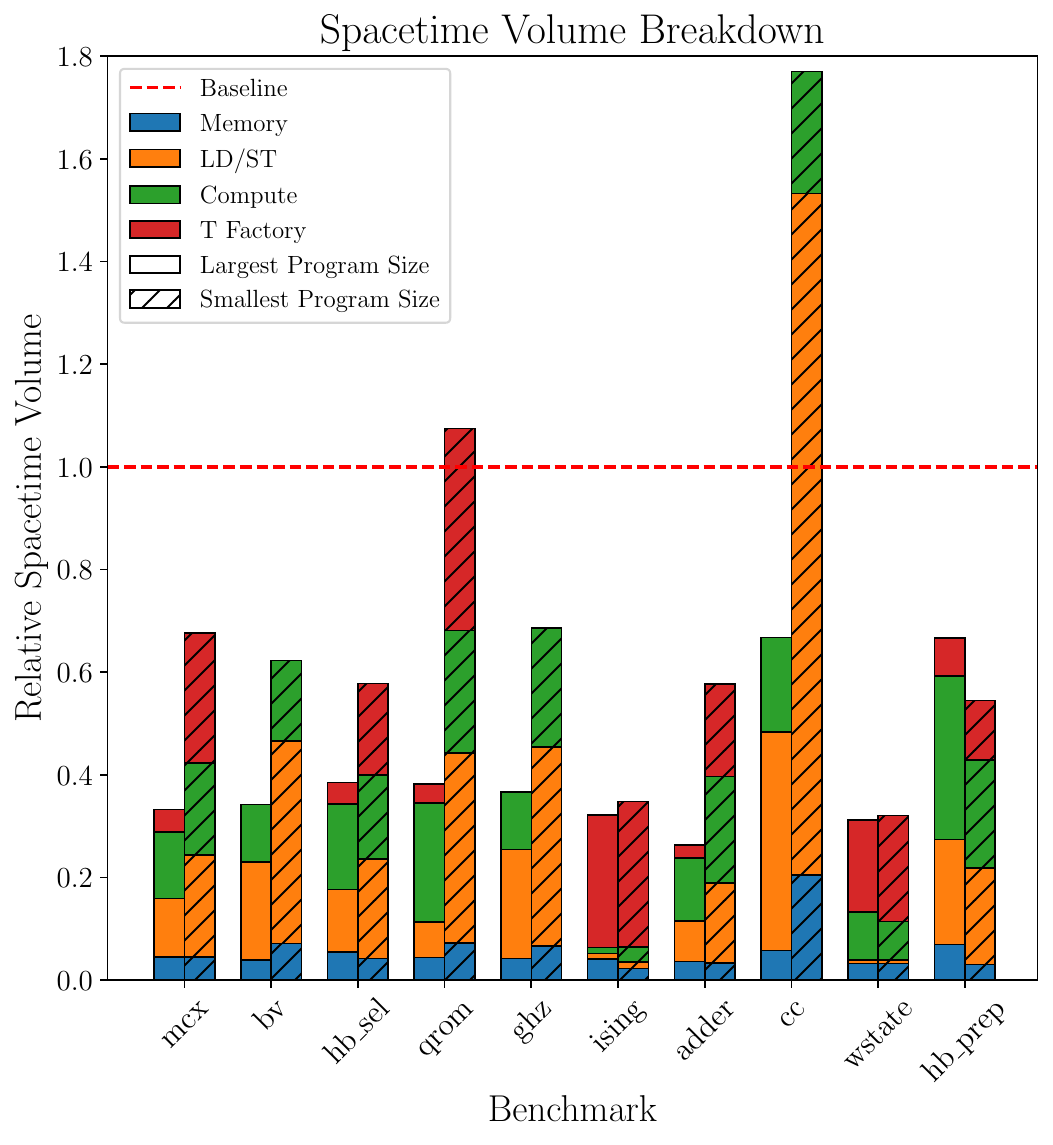}
    \caption{Bar plots denoting the breakdown of spacetime volume. A time step's load/store volume includes the compute space if all compute operations are waiting on loads/stores. This does not change the overall volume, but better highlights the cost of load/store.}
    \label{fig:breakdown}
\end{figure}

\begin{figure*}[t!]
    \centering
    \includegraphics[width=\linewidth]{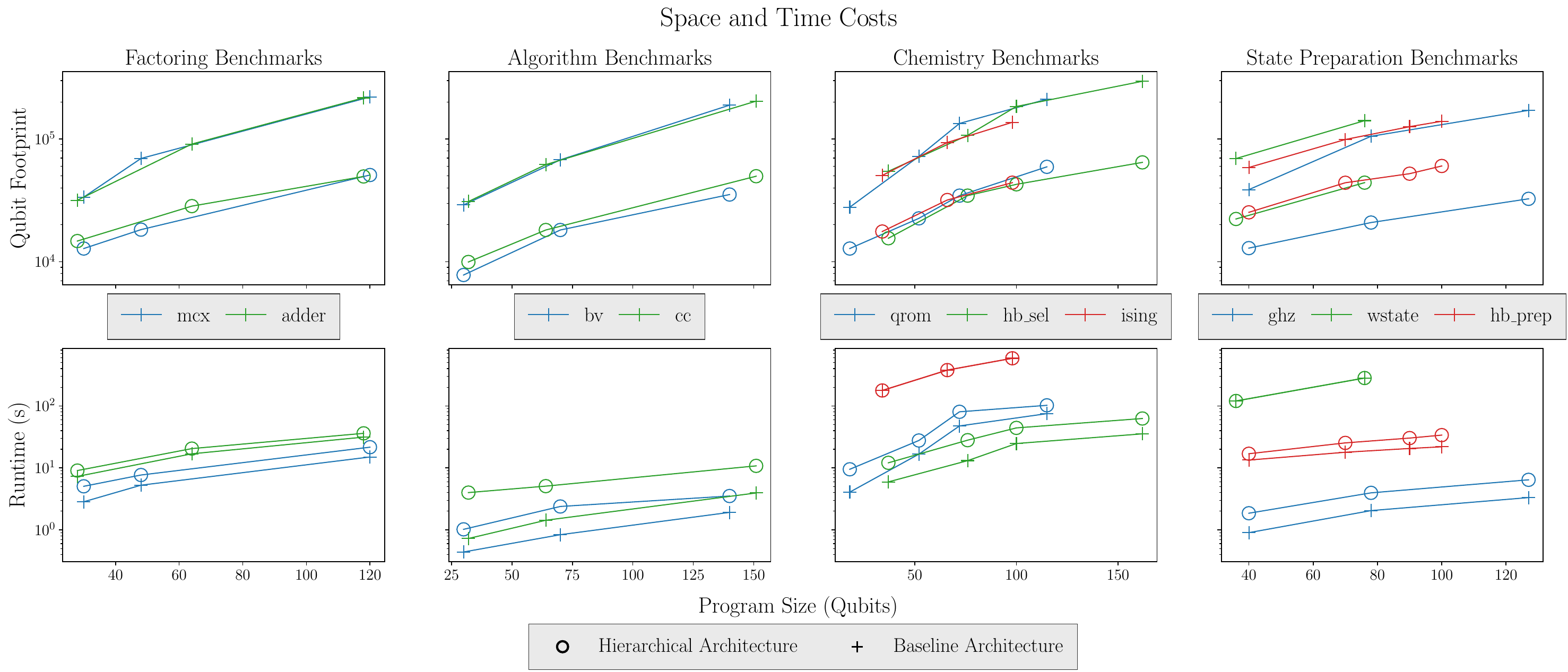}
    \caption{Comparison of the space (qubit footprint) and time (total runtime) costs to implement benchmark programs on a qSIEVE-powered hierarchical architecture against the surface code-only baseline architecture. Lower is better. The runtimes of two benchmarks, ising and wstate, are indistinguishable from the baseline. }
    \label{fig:space_time_cost}
\end{figure*}

\begin{figure*}[t!]
    \centering
    \includegraphics[width=\linewidth]{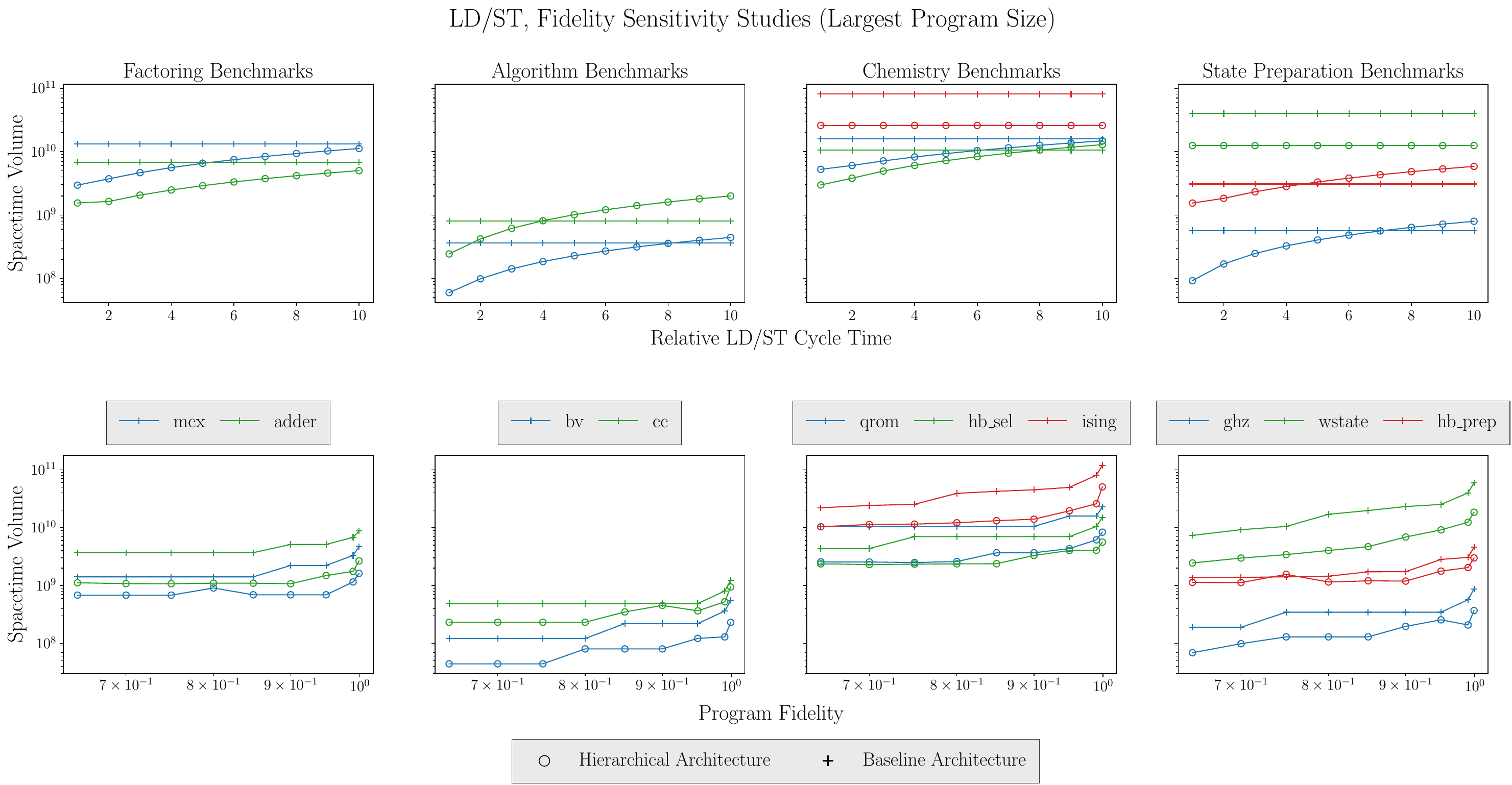}
    \caption{Line plots showing the sensitivity of the spacetime volume (qubit-seconds) to LD/ST cycle times relative to surface code cycle times (top) and required output fidelity (bottom), against the surface code-only baseline architecture. Lower is better.}
    \label{fig:sensitivity}
\end{figure*}

\begin{figure*}[t!]
    \centering
    \includegraphics[width=\linewidth]{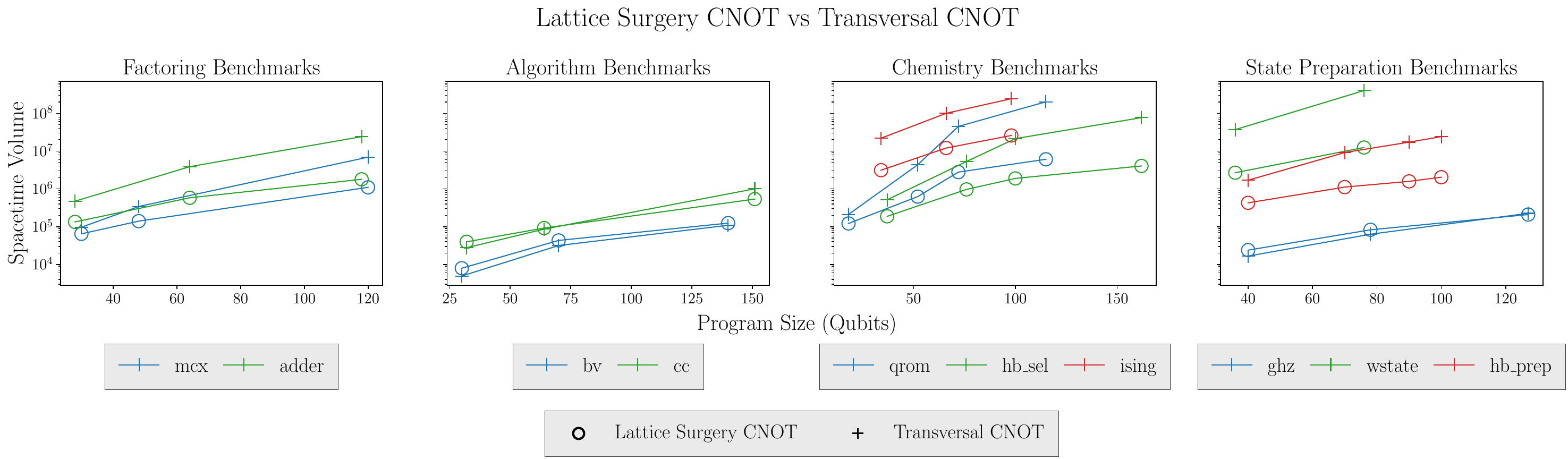}
    \caption{Comparison of spacetime volume (qubit-seconds) to implement benchmark programs using two methods implementing CNOT gates. Lower is better.}
    \label{fig:transversal}
\end{figure*}

\begin{figure*}[t!]
    \centering
    \includegraphics[width=\linewidth]{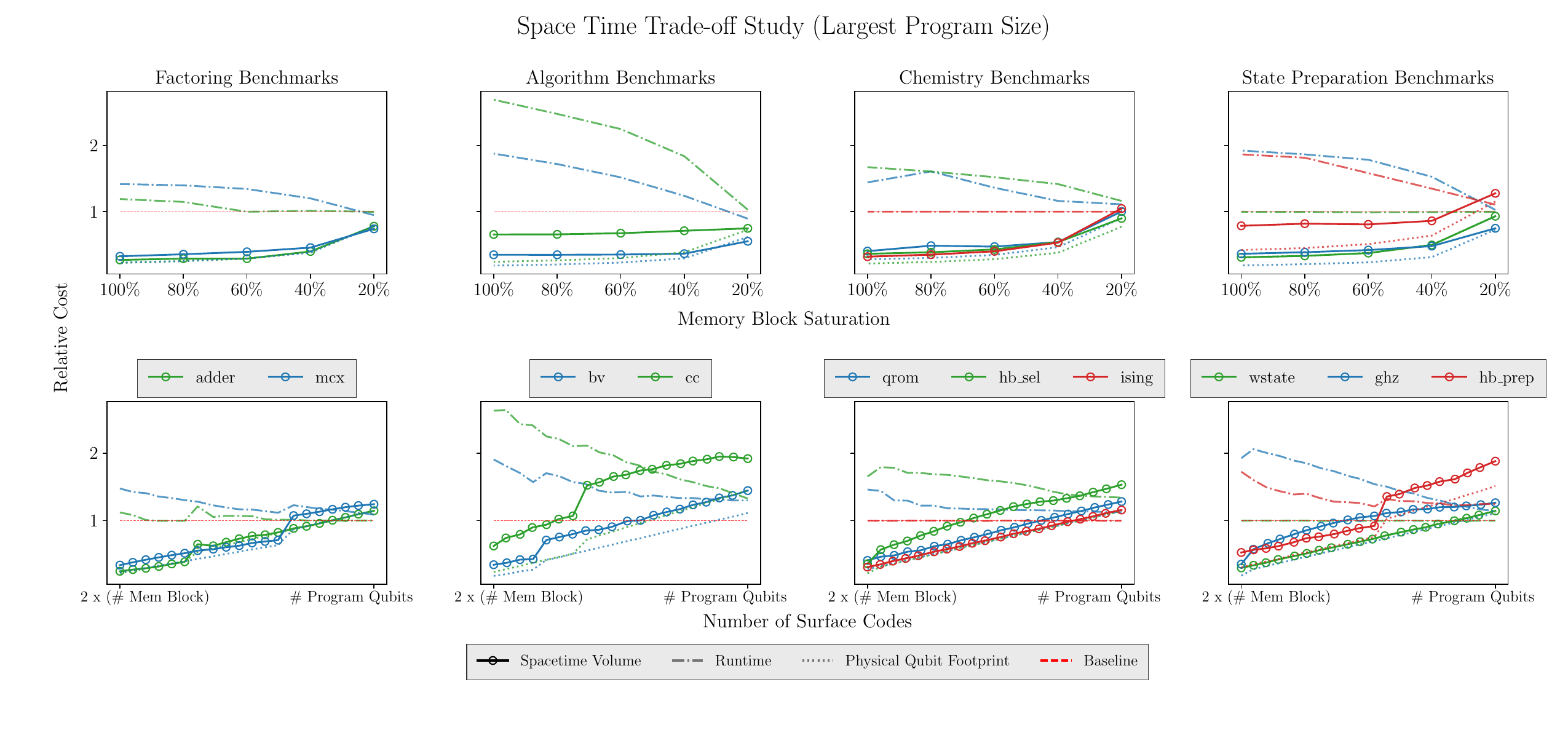}
    \caption{Line plots showing the (relative) space cost, time cost, and spacetime volume to implement benchmark programs as we decrease memory block saturation (increase the number of memory blocks, top) or increase the number of surface codes in the compute region (bottom). Lower is better. All costs are normalized by the surface code-only baseline. The resulting space-time trade-off can be seen as runtime decreases and qubit footprint increases with growing numbers of surface codes and memory blocks.}
    \label{fig:space_time_tradeoff}
\end{figure*}

\subsection{Benchmark Programs}\label{ssec:ft-benchmark}
To evaluate the performance of the hierarchical architecture, we choose a suite of available benchmark circuits that have diverse program features and applications, as summarized in Table~\ref{tab:benchcircs}. The main sources of benchmark circuits are open-source repositories qasmBench \cite{li2023qasmbench} and Cirq-ft \cite{cirq_developers_2023_8161252}. We identify four main areas of benchmark applications: factoring subroutines,  chemistry simulation subroutines, resource state preparation, and algorithms that demonstrate complexity theoretic quantum advantage. We also identify two key program features: serialization, which refers to the average ratio of inactive to active qubits in each time step; and T consumption level, which refers to the ratio of T gates to Clifford gates.

\subsection{Results and Discussions}\label{ssec:ft-results}
In this section, we summarize the performance results of the hierarchical architecture on the benchmark programs. All simulations assumed a physical error rate of $10^{-3}$. We note that our results are consistent with similar studies on qLDPC + surface code architectures~\cite{stein2025hetec}, further justifying the utility of qSIEVE. Our evaluation is necessarily distinct, however, since qSIEVE is a solution for reconfigurable atom arrays where CNOT gates in surface code have more than one implementation option and cycle times for qLDPC memory, LD/ST, and surface codes vary due to differences in movement costs.

Firstly, we demonstrate the advantage of the hierarchical architecture with generalized-bicycle codes by evaluating the spacetime costs of benchmark programs, compared against the cost in the surface code-only baseline architecture. The spacetime cost breakdowns are reported in Figure~\ref{fig:breakdown} and Figure~\ref{fig:space_time_cost}. Notably, the advantage of the hierarchical architecture is most significant when the benchmark programs have a high level of T-consumption and serialization. In the former, the program runtime is bottlenecked by magic state distillation, rendering the additional LD/ST costs negligible, given effective optimizations for qubit reuse.  Indeed, Figure~\ref{fig:breakdown} shows that for the benchmarks with a high level of T consumption, Ising model simulation and W-state preparation, spacetime volume is dominated by the need for a large T factory. As a result, as shown in Figure~\ref{fig:space_time_cost}, their runtimes on the hierarchical architecture match the baseline runtime with a significantly smaller qubit footprint due to efficient quantum memory, leading to reduced spacetime volumes. In a program with a high level of serialization, a large proportion of idle qubits benefit from the efficient, compressed memory. Furthermore, LD/ST time can be reduced via effective prefetching. Indeed, the benchmark program with the worst performance is the counterfeit coin algorithm, which had a low level of T consumption and a low level of serialization.

Secondly, we study the sensitivity of spacetime costs to key features of the hierarchical architecture, such as LD/ST time, required output fidelity, and implementation methods of CNOT gates. These studies are reported in Figure~\ref{fig:sensitivity} and Figure\ref{fig:transversal}. We begin by commenting on the robustness of the hierarchical architecture's advantage. First, we note that the sensitivity to required output fidelity is similar in both architectures due to the similar error suppression capabilities between the generalized-bicycle codes and surface codes. Secondly, as previously noted, variations in LD/ST costs have negligible effects on benchmark programs with a high T consumption level. In both cases, as shown in Figure~\ref{fig:sensitivity}, the advantage of a quantum memory hierarchy remains robust against hardware variations. Additionally, we note that for most benchmarks, a lattice surgery CNOT has lower spacetime costs due to ease of parallelism and movement-free routing. However, serial benchmarks with low T consumption, such as Bernstein-Vazarani and GHZ, perform better with a serial, movement-based transversal CNOT since they experience no slowdown and can instead benefit from the reduced space costs after removing routing surface codes. 

Finally, we explore the impact of the number of surface codes and the saturation of qLDPC memory (which decides the number of memory blocks). These define the overall qubit footprint and varying them enables a useful study of space-time trade-offs in the hierarchical architecture. In Figure~\ref{fig:space_time_tradeoff} we find that a hierarchical architecture with generalized-bicycle codes affords considerable freedom to balance time and space resources while maintaining an advantage over the surface code baseline. As we increase the number of memory blocks and surface codes, pressure on routing parallel CNOTs and LD/STs decreases, allowing for reduced runtimes. However, if optimizing for a balanced spacetime volume we find a high memory block saturation and few surface codes lead to the lowest overall costs due to the high qubit footprint of QEC codes.

\section{Conclusion}
In this work we presented qSIEVE, a protocol for implementing a restricted set of good non-local QEC codes in reconfigurable atom arrays. qSIEVE enables logical quantum memories with up to 10x fewer physical qubits than an equivalent memory using surface codes. Additionally, measuring a round of stabilizers with qSIEVE is 5-11x faster than alternative protocols for atom arrays, and we present a tiled memory design with shared controls to allow qSIEVE to scale in size effectively. 

Through detailed compilation and evaluation of key fault-tolerant benchmarks, we show the restricted set of codes enabled by our protocol is sufficient for a quantum memory hierarchy to outperform a standard surface code only architecture under a broad range of potential hardware costs. 

We hope this work motivates future research on hardware-tailored protocols for memory-efficient QEC codes as well as compilers for quantum memory hierarchies, which we believe will be critical
to organize the currently growing physical systems into
efficient, fault-tolerant systems capable of large-scale quantum
algorithms.

\section*{Acknowledgements}

This work is funded in part by the STAQ project under award NSF Phy-232580; in part by the US Department of Energy Office of Advanced Scientific Computing Research, Accelerated 
Research for Quantum Computing Program; and in part by the NSF Quantum Leap Challenge Institute for Hybrid Quantum Architectures and Networks (NSF Award 2016136), in part based upon work supported by the U.S. Department of Energy, Office of Science, National Quantum 
Information Science Research Centers, and in part by the Army Research Office under Grant Number W911NF-23-1-0077. The views and conclusions contained in this document are those of the authors and should not be interpreted as representing the official policies, either expressed or implied, of the U.S. Government. The U.S. Government is authorized to reproduce and distribute reprints for Government purposes notwithstanding any copyright notation herein. FTC is the Chief Scientist for Quantum Software at Infleqtion and an advisor to Quantum Circuits, Inc. JL is supported in part by the University of Pittsburgh, School of Computing and Information, Department of Computer Science, Pitt Cyber, PQI Community Collaboration Awards, John C. Mascaro Faculty Scholar in Sustainability, NASA under award number 80NSSC25M7057, and Fluor Marine Propulsion LLC (U.S. Naval Nuclear Laboratory) under award number 140449-R08. This research used resources of the Oak Ridge Leadership Computing Facility, which is a DOE Office of Science User Facility supported under Contract DE-AC05-00OR22725.



\bibliographystyle{ACM-Reference-Format}
\bibliography{refs}

\end{document}